 \documentclass[11pt,twocolumn]{article}

 \pdfoutput=1

\usepackage[round]{natbib}   
\usepackage[english]{babel}
\usepackage{amsmath}
\usepackage{enumitem}
\usepackage{color}
\usepackage{float}
\usepackage{ifthen}
\usepackage{hyperref}

\usepackage{titlesec}
\titlespacing*{\section}{0pt}{1.1\baselineskip}{\baselineskip}

\usepackage{amsmath,amsfonts,amssymb,bm}

\usepackage[normalem]{ulem}
\usepackage{graphicx,siunitx,xspace}
\usepackage{booktabs}

\usepackage{chemformula}
\usepackage{epstopdf}

\newcommand{\cpd}{c_{p_\mathrm{d}}}
\newcommand{\cpv}{c_{p_\mathrm{v}}}
\newcommand{\cpe}{c_\mathrm{e}}

\newcommand{\cpl}{c_{\mathrm{l}}}
\newcommand{\cpell}{c_\ell}

\newcommand{\rr}{\mathrm r}

\newcommand{\rx}{\mathrm x}

\newcommand{\re}{\mathrm{e}}

\newcommand{\rs}{\ast}

\newcommand{\rd}{\mathrm d}
\newcommand{\rt}{\mathrm t}

\newcommand{\rv}{\mathrm v}
\newcommand{\rl}{\mathrm l}

\newcommand{\ri}{\mathrm i}
\newcommand{\EURECA}{EUREC$^4\!$A\xspace}
\newcommand{\EC}{\EURECA-Circle\xspace}

\begin{document}

\def\hmath#1{\text{\scalebox{1.5}{$#1$}}}
\def\lmath#1{\text{\scalebox{1.4}{$#1$}}}
\def\mmath#1{\text{\scalebox{1.2}{$#1$}}}
\def\smath#1{\text{\scalebox{.8}{$#1$}}}

\def\hfrac#1#2{\hmath{\frac{#1}{#2}}}
\def\lfrac#1#2{\lmath{\frac{#1}{#2}}}
\def\mfrac#1#2{\mmath{\frac{#1}{#2}}}
\def\sfrac#1#2{\smath{\frac{#1}{#2}}}

\def\pow{^\mmath}

\twocolumn[

 \begin{center}
{\bf \Large {
 On moist potential temperatures and their ability to characterize 
\\ \vspace*{2mm} 
 differences in the properties of air parcels}}
\\
\vspace*{2mm}
{\Large by Pascal Marquet$\:{}^{(1)}$
       and Bjorn  Stevens$\:{}^{(2)}$}.
\\
\vspace*{2mm}
{\large ${}^{(1)}$ CNRM, Universit\'e de Toulouse, 
        M\'et\'eo-France, CNRS, Toulouse, France.}
 \\
\vspace*{1mm}
{\large ${}^{(2)}$ Max Planck Institute for 
Meteorology, Hamburg, Germany.
}
\\
\vspace*{1mm}
{bjorn.stevens@mpimet.mpg.de \\
pascal.marquet@meteo.fr / 
pascalmarquet@yahoo.com}
\\
\vspace*{1mm}
{Submitted first to the 
{\it Journal of Atmospheric Sciences\/} on the 2nd of April, 2021.}\\
{Fist revised version: 17th of August, 2021.}\\
{Second revised version: 23th of November, 2021.}
 \end{center}

%

\begin{center}
{\large \bf Abstract}
\end{center}
\vspace*{-3mm}

\hspace*{7mm}
A framework is introduced to compare moist `potential' temperatures. 
The equivalent potential temperature, $\theta_\re,$ the liquid water potential temperature, $\theta_\ell,$ and the entropy potential temperature, $\theta_s$, are all shown to be potential temperatures, in the sense that they measure the temperatures of certain reference state systems whose entropy is the same as that of the air-parcel. They only differ in the choice of reference state composition: $\theta_\ell$ describes the temperature a condensate-free state, $\theta_\re$ a vapor-free state, and $\theta_s$ a water-free state would require to have the same entropy as the given state. Although in this sense $\theta_\re,$ $\theta_\ell,$ and $\theta_s$ are all different flavors of the same thing, only $\theta_\ell$ satisfies the stricter definition of a `potential temperature', as corresponding to a reference temperature accessible by an isentropic and closed transformation of a system in equilibrium; both $\theta_\re$ and $\theta_\ell$ measure the `relative' enthalpy of an air parcel at their respective reference states; but only $\theta_s$ measures air-parcel entropy. None mix linearly, but all do so approximately, and all reduce to the dry potential temperature, $\theta$ in the limit as the water mass fraction goes to zero. As is well known, $\theta$ does mix linearly and inherits all the favorable (entropic, enthalpic, and potential temperature) properties of its various -- but descriptively less rich -- moist counterparts. All, involve quite complex expressions, but admit relatively simple and useful approximations. Of the three moist `potential' temperatures, $\theta_s$ is the least familiar, but the most well mixed in the broader tropics, a property that merits further study as a possible basis for constraining mixing processes.
\vspace*{6mm}
%
] 

\section{Introduction}\label{sec:Introduction} 
\vspace*{-3mm}

The strong pressure dependence of many state variables can complicate attempts to compare the properties of atmospheric air parcels. For dry air, approximated as an ideal gas, the potential temperature, $\theta,$ elegantly describes an air parcel's state. It does so by accounting for the effect of pressure on the state of the air parcel, which then facilitates comparisons of the properties of air-parcels independent of their ambient pressure.

Physically, $\theta$ describes the temperature dry air would attain, were it brought adiabatically to a reference pressure, usually (but not necessarily) taken to be the standard pressure, $P_0 = \SI{1000}{\hecto\pascal}$. Mathematically
\vspace*{-3mm}
\begin{equation}
 \theta = T \left(\frac{P_0}{P}\right)^{\kappa_\rd},
\end{equation}
where $T$ is temperature (in Kelvin), $P$ pressure, and $\kappa_\rd = R_\rd/\cpd$.
Values of the dry-air gas constant, $R_\rd,$ and the isobaric specific heat, $c_{p_\rd},$ depend on how dry-air is defined --- usually as \ch{N2}, \ch{O2}, \ch{Ar}, and sometimes \ch{CO2}, approximated as an ideal mixture of ideal gases specified in some fixed proportion.\footnote{For the treatment of the thermodynamics it is common to neglect changing \ch{CO2} and \ch{O2} from burning fossil fuels.} As has been appreciated for some time \citep{Bezold_1888a,Bauer_1910}, $\theta$ is much more than a way to compensate for the effects of pressure on temperature, it measures the buoyancy of air parcels (on isobars), the enthalpy of the air at a reference pressure, and it mixes linearly. It is also linearly proportional to the exponential of the dry air-parcel entropy, $s$ divided by $\cpd$.

For a variable composition fluid, even for the limiting case of an ideal mixture of ideal gases, the situation is more complicated. Admitting condensible phases for the minor constituent, what we call moist-air, complicates matters further. Earth's atmosphere is, however, fundamentally comprised of moist-air -- it cannot be understood without considering the water it contains. This makes it necessary to address these complications, and explains the rich literature that has developed, proposing one or the other generalization of the idea of the potential temperature to moist-air. As it turns out, these \textit{moist potential temperatures} all measure slightly different quantities, and while this point is generally well understood \cite[see][]{Pauluis2018}, physical understanding of exactly what they measure remains rudimentary.

\begin{figure}[htb]
\centering
\includegraphics[width=0.85\linewidth]{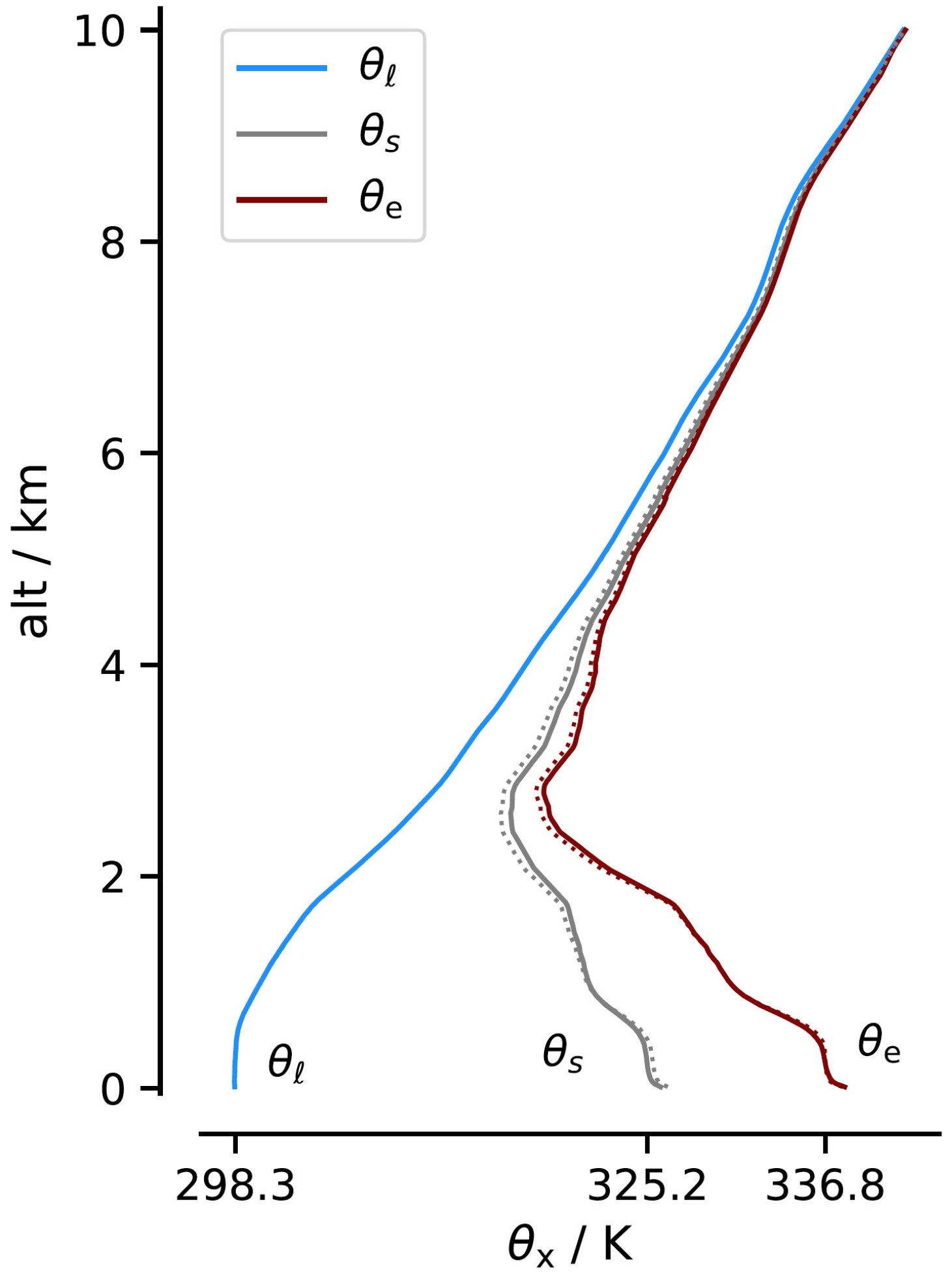}
\vspace*{-3mm}
\caption{\it \small 
Mean profiles of the liquid water, $\theta_\mathrm{l},$ entropy, $\theta_s,$ equivalent $\theta_\mathrm{e}$ potential temperatures from 757 dropsondes launched along the \EC \citep{StevensEtAl2021,GeorgeEtAl2021}. 
Dashed lines are from slight approximations to the same quantities, as discussed later in the manuscript.} \label{fig:dropsondes}
\end{figure}

Given the prominence of a literature that has made statements to the contrary \cite[cf.,][]{Emanuel1994,PauluisEtAl2008,Raymond2013,Romps2015}, it may come as a surprise that the equivalent potential temperature, $\theta_\re,$ does not measure the entropy of an air-parcel -- not even approximately. Fig.~\ref{fig:dropsondes}, which presents vertical profiles of the liquid-water potential temperature, $\theta_\ell,$ the entropy potential temperature, $\theta_s,$ and $\theta_\re$ as calculated from thermodynamic measurements made during the recent \EURECA field study, substantiates this point. All quantities are invariant for isentropic transformations of closed air parcels (namely with a constant total water content), but this adiabatic invariance, as the figure demonstrates, does not mean that their differences (e.g., with altitude) are indicative of differences in entropy, nor does it guarantee that their isopleths are isentropes. Were this the case, then $\theta_\re,$ or $\theta_\ell$ could not take on different values for the same value of $\theta_s.$ 

As it turns out, only $\theta_s \propto \exp(s/c_{p_\rd})$, and hence only its differences measure differences in $s.$ Why this is so, how $\theta_s$ and the other moist potential temperatures relate to one another, and hence what precisely one compares when one compares their different values, are the subject of this paper. We begin by constructing a framework (\S\ref{sec:preliminaries}) that allows us to define precisely what we mean by the term potential temperature. This framework is used in \S\ref{sec:moist-potential-temperatures} to derive exact (within the framework of the given assumptions) expressions for $\theta_\ell,$ $\theta_s,$ and $\theta_\re.$ These are physically interpreted, and compared to common simplified expressions of the same variables. The ability of the different moist potential temperatures to measure different air-parcel properties is evaluated in \S\ref{sec:properties}. In \S\ref{sec:Examples} examples are chosen to substantiate some of our main points, before concluding in \S\ref{sec:conclusions}.

\section{Terminology and Definitions} 
\label{sec:preliminaries}
\vspace*{-4mm}

\subsection{Moist-air} 
\label{sec:preliminaries_moist_air}
\vspace*{-4mm}

We idealize the atmosphere as \textit{moist-air}, i.e., as a mixture of dry air and water, allowing a portion of the latter to condense as conditions dictate. In equilibrium, the thermodynamic state of the moist-air is completely specified by three thermodynamic coordinates. For these we adopt the temperature, $T,$ the pressure $P$ and the water mass fraction (total-water specific humidity), $q_\rt .$ A guide to the subscript notation adopted is given in Table~\ref{tab:notation}.

\begin{table*}[t]
\caption{\it Subscript notation for the specification of particular states.}\label{tab:notation}
\vspace*{-2mm}
\begin{center}
\begin{tabular}{ll}
 \hline \hline
 Subscript & Description \\ \midrule
 0 & standard values, or quantities evaluated at standard values \\
 r & a non-standard reference value \\
 1, 2 & to distinguish different values or states, indexed by $j$ \\
 v, l, t & vapor, liquid and total water \\
 $\ast$ & value at vapor-liquid (water) saturation \\
 $\rd$ & dry component ($q_\rd = 1-q_\rt$ for the two components system) \\
 $\re$ & equivalent (liquid) reference state ($q_\rl = q_\rt$) \\
 $\ell$ & liquid-free reference state ($q_\rv = q_\rt$) \\ 
 $s$ & absolute entropy value \\
 $\rx$ & unspecified reference state defining the moist potential temperature, $\theta_\rx$ \\
 \hline
\end{tabular}
\end{center}
\vspace*{-6mm}
\end{table*}

To arrive at an analytically tractable description, and to facilitate precise statements, we make four further assumptions: 
(i) the specific heats are approximated as constant, i.e., not varying with temperature, 
(ii) the non-condensate phase (gas/vapor) is approximated to behave as an ideal mixture of ideal gases; 
(iii) the contribution of the condensate to the total volume is negligible; and 
(iv) only a single condensate phase is admitted, and this is treated as an ideal liquid, whose mass fraction is denoted $q_\rl.$

Assumptions (i) to (iii) are a common starting point for atmospheric thermodynamics \citep{Emanuel1994,stevenssiebesma2020,Romps2021}, which facilitates analytic work \citep{Ooyama1990,Raymond2013,Romps2017}. Approximation (iv) is adopted because including the ice phase introduces formal complexity that is not relevant to our arguments. Approximations (i) and (ii) can be relaxed by using the variable values of specific heat and non-ideal effects based on IAPWS and TEOS10 tools \citep{IAPWS10,Feistel_2018}, but sacrifices analytic clarity for accuracy.
\vspace*{-4mm}

\subsection{Reference states and notation} 
\label{sec:preliminaries_notations}
\vspace*{-3mm}

Many thermodynamic state functions, such as the entropy or the enthalpy are defined with respect to some reference state value. For ice-free moist-air in thermal equilibrium, a reference state can be fully characterized by specifying a reference temperature, $T_\rr,$ and the reference state composition $\{P_{\rd,\rr},q_{\rv,\rr},q_{\rl,\rr}\}.$ Here $P_\rd$ denotes the partial pressure of the dry-air, and the roman-subscript '$\rr$' denotes a reference value. A description in terms of three (rather than two) additional state variables anticipates the possibility of mechanical disequilibrium\footnote{What we call mechanical equilibrium, which is a force (pressure) balance between phases, is sometimes referred to as phase equilibrium.}. This possibility is required to accommodate the derivation of some of the moist potential temperatures in the proposed framework.

The specification of the reference state already illustrates how notation can be a challenge. In the present manuscript, subscripts are used to give specificity to a class of variables. For instance, roman subscript $\rd,$ $\rv,$ and $\rl$ are used to distinguish properties of dry-air versus water-vapor or liquid-water. Subscript $\rt$ is used to denote total water, whereby in an ice-free system $q_\rt = q_\rv + q_\rl.$ In addition, we introduce the roman subscript $\rr$ to identify a reference state value, and $\rx$ to denote quantities associated with a particular choice of reference-state composition. As a rule, and as summarized in Table~\ref{tab:notation}, numeric subscripts are used to distinguish different air-parcel states (with 0 denoting standard values), and letters are used to denote a particular disposition of matter.

Three special compositions of the reference state are defined as special cases of $\rx.$ These correspond to end-member (or limiting) situations whereby:
\vspace*{-3.5mm}
\begin{description}
\item[e state:] denotes, the `equivalent' composition of the reference state, whereby $\rx \rightarrow \{P_{\rd,0},0,q_\rt\},$ and hence is vapor free; 
\vspace*{-3mm}
\item[$\mathbf{\ell}$ state:] denotes, the `liquid-less' composition of the reference state, whereby $\rx \rightarrow \{P_{\rd,0},q_\rt,0\},$ and hence is condensate free;
\vspace*{-3mm}
\item[$\mathbf{s}$ state:] denotes, the `entropic' composition of the reference state, whereby $\rx \rightarrow \{P_{\rd,0,}0,0\},$ and hence is water-free (dry). 
\end{description}
\vspace*{-3mm}
Dalton's law and state equations specifiy the partial pressures of the ideal gases in terms of the total pressure and the specific humidities, such that
\vspace*{-2mm}
\begin{equation}
P_{\rd,0} \equiv P_0 \left(\frac{1-q_{\rt,\rr}}{1+\varepsilon \:q_{\rv,\rr} - q_{\rl,\rr}} \right), 
\;\; 
\text{with} 
\;\; 
\varepsilon = \frac{R_\rv}{R_\rd} -1 \:, \label{eq:Pd0} 
\end{equation}
and correspondingly for $P_{\rv,0} = P_\rr - P_{\rd,0}.$ The names (equivalent, liquid-less, entropic) are not especially informative nor intuitive in this context, but chosen to keep consistency with historical usage. 
\vspace*{-3mm}

\subsection{The entropy temperature, $\vartheta$} 
\label{sec:preliminaries_T}
\vspace*{-3mm}

For an ideal gas the specific entropy is given as 
\vspace*{-2mm}
\begin{equation}
s(T,P) \: = \: s(T_\rr,P_\rr) 
+ c_p\ln\!\left(\frac{T}{T_\rr}\right) 
- R \: \ln\!\left(\frac{P}{P_\rr}\right) \: ,
\label{eq:entropy}
\end{equation}
\vspace*{-7mm}

\noindent
where $s(T_\rr,P_\rr),$ alternatively written as $s_\rr,$ is the entropy at the reference temperature, $T_\rr,$ pressure, $P_\rr,$ and gas constant $R$ \citep{Fermi1937}. The entropy for the condensed phase can be expressed similarly, except that the dependence on pressure vanishes by virtue of the assumed incompressibilty. The sum of the constituent entropies, defines the system entropy, 
\vspace*{-3mm}
\begin{equation}
s \; = \; 
        q_\rd \: s_\rd 
\: + \: q_\rv \: s_\rv 
\: + \: q_\rl \: s_\rl \: . 
\label{eq:ssum}
\end{equation}
\vspace*{-8mm}

Combining Eqs.~(\ref{eq:entropy}) with (\ref{eq:ssum}) results in a long expression for the entropy \cite[see for instance][where however the dry-air and liquid-water or water-vapour reference entropies are sometimes set to zero]{Pauluis_Czaja_Korty_2010,Marquet2011QJ,Raymond2013,stevenssiebesma2020}.
By referring the component systems to a common reference temperature, the remaining terms can be combined into a single function, $\vartheta,$ such that without any further loss in generality:
\vspace*{-2mm}
\begin{equation}
s(T, P, q_\rt) - s_\rr 
\: = \: c_{p,\rr} \: \ln \!\left( \frac{\vartheta}{T_\rr} \right),
\label{eq:s}
\end{equation}
\vspace*{-6mm}

\noindent
where
$s_\rr \equiv s(T_\rr,P_\rr, q_{\rv,\rr},q_{\rl,\rr}),$ with $P_\rr = P_{\rd,\rr} + P_{\rv,\rr},$ and $c_{p,\rr}$ is the isobaric specific heat of the system in the reference state. Expanding Eq.~(\ref{eq:ssum}) shows $\vartheta$ to be a function of both the state $(T,P,q_\rt)$ and those aspects of the reference state that determine its composition, i.e., $\{P_\rr, q_{\rv,\rr}, q_{\rl,\rr}\}.$ 

\vspace*{-1mm} \noindent
Using the subscript $\rx$ to remind us of how quantities depend on the composition of the reference state, allows us to write Eq.~(\ref{eq:s}) as
\vspace*{-3mm}
\begin{equation}
s(T, P, q_\rt) - s_\rx(T_\rr) 
\; = \; c_\rx \:
\ln \! \left( 
\frac{\vartheta_\rx(T,P,q_\rt)}{T_\rr} 
\right) \:.
\label{eq:sx}
\end{equation}

\vspace*{-3.5mm}
\noindent
Readers familiar with the \cite{HaufHoeller1987} entropy temperature will recognize it as the special case of $\vartheta_\rx$ with $\rx=\re,$ i.e., the vapor-free reference state {$\{P_\rr=P_0, q_{\rv,\rr}=0,q_{\rl,\rr}=q_\rt\}.$} Leaving the composition of the reference state open (as indicated by $\rx$), enables a more general treatment of the moist potential temperatures $\vartheta_\rx.$ Notwithstanding this generalization, and the fact that $\vartheta_\rx$ measures entropy differences ($s-s_\rx(T_\rr)$) rather than the entropy, $s,$ itself, we also refer to $\vartheta_\rx$ as an `entropy temperature'. Calling it a `generalized entropy-difference temperature' would be more precise, but unwieldy and ahistorical.
\vspace*{-3mm}

\subsection{Potential temperatures} 
\vspace*{-3mm}

Based on Eq.~(\ref{eq:sx}) we formalize the \cite{stevenssiebesma2020} definition of potential temperature as follows
\vspace*{-3mm}
\begin{quote}
Given a reference state whose \textit{composition} is denoted by $\rx,$ the potential temperature $\theta_\rx$ is the temperature ($T_\rr$) this reference state must adopt to have the same entropy as the given state. 
\end{quote}

\vspace*{-2mm} \noindent
Mathematically this defines $\theta_\rx,$ implicitly to satisfy $s - s_\rx(\theta_\rx) = 0,$ for some given specification of $\rx$.
It follows that $\theta_\rx = \vartheta_\rx.$ The adjective `potential' describes how $\theta_\rx$ is the temperature the system \textit{would} adopt were it brought to the reference state without changing its entropy. The potential temperature as defined above is thus a generalization of the \cite{HaufHoeller1987} entropy temperature.

By definition, $\theta_\rx$ is invariant for any isentropic transformation that does not imply a change in the reference state composition. For dry air, the reference state is completely specified by $P_\rr$ the reference state pressure, usually taken to be standard pressure, $P_0.$ A stricter form of the above definition, and one satisfied by the dry-air potential temperature, $\theta,$ would additionally require the transformations to be closed and reversible, but by this definition there can be at most one moist potential temperature. 

\begin{figure*}[htb]
\centering\noindent
\includegraphics[width=0.99\linewidth]{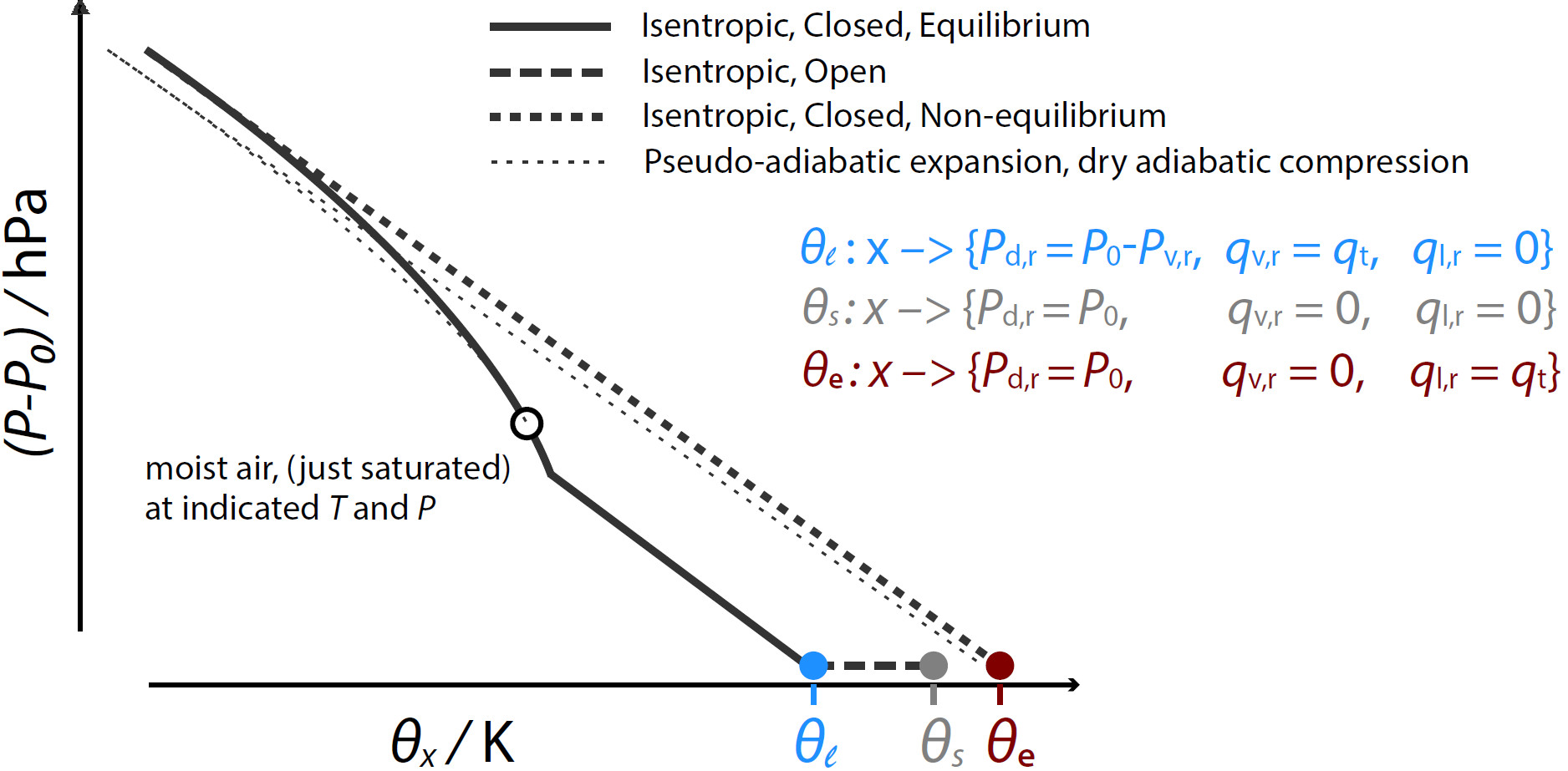}
\vspace*{-3mm}
\caption{\it \small 
Schematic of the moist potential temperatures as projected in 
$(\theta_\rx,P)$ coordinates. All connected points share the same entropy, with their different reference temperatures reflecting differences in the reference state composition. The types of isentropic transformations that connect the different moist potential temperatures are indicated in the legend. That $\theta_s$ is in position $2/3$ between $\theta_\ell$ and $\theta_e$ is explained further in the text.}
\label{fig:moist-thetas}
\end{figure*}

The three moist potential temperatures, $\theta_\ell,$ $\theta_s,$ and $\theta_e$ are shown below to correspond to the three limiting reference states ($\re,\ell,s$) described above. Each is illustrated schematically in Fig.~\ref{fig:moist-thetas}, whereby all of the points connected by lines in the figure share the same entropy, but the transformations that bring them to their respective reference states differ. Reversible transformations of the closed system are shown along the solid line, and the dashed-dotted lines show either non-equilibrium transformations (for instance associated with the condensate for $\rx=\re$), or open transformations as associated with removing the water substance at constant entropy, for $\rx=s$
\vspace*{-3mm}

\section{Moist potential temperatures} 
\label{sec:moist-potential-temperatures}
\vspace*{-3mm}

\subsection{The equivalent potential temperature, $\theta_\re$} 
\label{--Subsection:Thetae}
\vspace*{-2mm}

The oldest, and most familiar, moist potential temperature, $\theta_\re,$ was introduced by \cite{Rossby1932} as the value of $\theta$ for a parcel undergoing an infinite pseudo-adiabatic ascent toward $P=0$, with all the water removed by precipitation (Fig.~\ref{fig:moist-thetas}). Hence it measures the potential temperature required of dry air, such that following an adiabatic expansion its temperature asymptotically approaches that of moist air expanded pseudo-adibatically -- herein lies the modern idea of equivalence.\footnote{\cite{MarquetThibaut2018JAS} traces the idea of an `equivalent' potential temperature, to \cite{Normand21} who introduced it as a generalization of \citet[p.18]{Schubert1904} and \citet[p.3]{Knoche1906}, and ultimately von Bezold's concept of (`higher' or `supplemented' or `complete') `equivalent' temperature ($T_e$). Normand's equivalent potential temperature ($\theta_\re$) was defined using $c_{p\rd} \: T_e = c_{p\rd} \: T + \ell_\rv \: q_\rv$ and $\theta_\re \approx \theta \: (T_e/T)$, where the $T_e$ was the `moist equivalent' of $T$ with the impact of $q_\rv$ taken into account via the moist `total energy'.}

In the present paper, and following contemporary usage, $\theta_\re$ is defined as the temperature of a vapor-free (equivalent) reference state, with the same entropy, but for which all the water is in the condensate phase at the standard pressure, $P_0.$ This differs from Rossby's definition by virtue of being isentropic (condensate is not precipitated from the parcel), hence `equivalence' is being drawn to a system in which the specific heat of the water mass is retained. Retaining the condensate maintains a closed system, but comes at the cost of the reference state being in a state of mechanical disequilibrium ($P_{\rv,\rr} \ne P_\rs(T_\rr)$).

In the equivalent (vaporless) reference state the specific heat and gas constants become 
\vspace*{-2.5mm}
\begin{equation}
\cpe = c_{p\rd} \:(1-q_\rt) + c_\rl \:q_\rt 
\quad \text{and} \quad 
R_\re = R_\rd \: (1-q_\rt).
\end{equation}
With $q_\rl = q_\rt - q_\rv,$ Eq.~(\ref{eq:ssum}) becomes
\vspace*{-1.5mm}
\begin{equation}
\!\!\!\!
s \, - \, s_\re(T_\rr) = 
\cpe \: \ln \!\left( \frac{T}{T_\rr} \right)
- \, R_\re \: \ln\! \left( \frac{P_\rd}{P_0} \right) 
+ \, q_\rv \: (s_\rv - s_\rl) \!
\label{eq:entropy-equiv}
\end{equation}

\vspace*{-3mm} \noindent
where
\vspace*{-3mm}
\begin{equation}
s_\re(T_\rr) 
\; = \;
(1-q_\rt) \: s_\rd(T_\rr,P_0) 
\: + \: q_\rt \: s_\rl(T_\rr),
\label{eq:sre} 
\end{equation}

\vspace*{-2mm} \noindent
is the reference entropy of the `equivalent' state at temperature $T_\rr$. The `equivalent' reference state, defined earlier as a dry-air and vapor-free fluid parcel with all water content in the liquid state, results from this Eq.~(\ref{eq:sre}) for $s_\re(T_\rr)$ which depends only on $s_\rd$ and $s_\rl$, as in \cite{HaufHoeller1987}. A possible physical process for realizing such a state would be to separate the condensate as it forms, maintaining it in condensate form in thermal equilibrium with the gas -- for instance by separating it from the gas with the help of a perfectly conducting membrane whose own heat capacity is vanishingly small.

At the temperature $T,$ the entropy difference between two phases can be written relative to the saturation entropy, $s_\rs$:
\vspace*{-2.5mm}
\begin{equation}
\!\!\!\!\!
s_\rv \, - \, s_\rl = 
(s_\rv - s_\rs) \, + \, (s_\rs - s_\rl) 
= -R_\rv 
\ln \!\left( \frac{P_\rv}{P_\rs}\right) 
+ \frac{\ell_\rv}{T},
\label{eq:svmsl} 
\end{equation}

\vspace*{-2.5mm} \noindent
where $P_\rs$ is the saturation vapor pressure and $\ell_\rv \; = h_\rv - h_\rl$ is the vaporization enthalpy (or latent heat of vaporization). Recasting the above in the form of Eq.~(\ref{eq:sx}), and defining $\theta_\re$ as the value of the reference temperature that satisfies $s_\re(\theta_\re) = s,$ yields:
\vspace*{-2mm}
\begin{equation}
\theta_\re \: = \: 
\vartheta_{\rx=\re} \: = \: 
T 
\: 
\left( \frac{P_0}{P}\right)^{\kappa_\re}
\:
\exp \! \left( \frac{q_\rv \: \ell_\rv}{\cpe \: T}\right)
\;\: \Omega_\re \: ,
\label{eq:thetae}
\end{equation}

\vspace*{-3mm} \noindent
where
\vspace*{-2mm}
\begin{equation}
\Omega_\re = \left(\frac{R}{R_\re}\right)^{\kappa_\re} \left(\frac{P_\rv}{P_\rs}\right)^{-q_\rv R_\rv/\cpe}
\label{eq:omegae}
\end{equation}
and $R = R_\rd \, (1-q_\rt) + R_\rv \, q_\rv$ is the moist-air gas constant. Eq.~(\ref{eq:thetae}) is identical to that found in modern textbooks, \cite[e.g.,][]{Emanuel1994,stevenssiebesma2020}. The factor, $\Omega_\re,$ which is close to $1,$ is often neglected in practical applications. Its first term, proportional to the ratio of gas constants, absorbs the effect of defining $\theta_\re$ in terms of $P$ rather than $P_\rd,$ and its exponent, $\kappa_\re = R_\re / \cpe,$ which also appears in Eq.~(\ref{eq:thetae}), generalizes the analogous quantity for dry air to the equivalent reference system. 

Physically, $\theta_\re$ measures the temperature air would have if all of its vaporization enthalpy were used to warm the parcel (accounting for the specific heat of the condensate) at standard pressure. It does not satisfy our stricter definition of a potential temperature as the reference state is not in mechanical equilibrium, and complete condensation through expansional cooling can only be realized asymptotically.
\vspace*{-2mm}

\subsection{The liquid-water potential temperature, $\theta_\ell$} 
\label{--Subsection:Thetal}
\vspace*{-2mm}

The liquid-water potential temperature, $\theta_\ell,$ was introduced by \cite{Betts1973} to study reversible changes of phase in non-precipitating shallow convection. Following \cite{stevenssiebesma2020} it can be derived from Eq.~(\ref{eq:sx}) by adopting a liquid-free reference state, denoted $\ell,$ in which all the condensate is assumed to be in the vapor phase at standard pressure and at temperature $T_{\rr,\ell}$. 
The corresponding specific heat and gas constants are
\vspace*{-2mm}
\begin{equation}
\!\!\!\!\!\!
\cpell = \cpd (1-q_\rt) + \cpv \:q_\rt,
\; \text{and} \; 
R_\ell = R_\rd (1-q_\rt) + R_\rv q_\rt,
\end{equation}

\vspace*{-3mm}\noindent
and the reference entropy becomes
\vspace*{-2mm}
\begin{equation}
s_\ell = (1-q_\rt)\: s_\rd(T_\rr,P_{\rd,0}) + q_\rt \: s_\rv(T_\rr,P_{\rv,0})
\label{eq:srl}
\end{equation}
where $P_{\rv,0} = P_0 - P_{\rd,0}$ with $P_{\rd,0}$ defined by Eq.~(\ref{eq:Pd0}). As the $\ell$-state is comprised of a dry-air and water vapor alone, a state of mechanical (phase) equilibrium is ensured as long as 
\vspace*{-2mm}
\begin{equation}
q_\rt < 
\frac{ R_\rd \, P_\rs(T_\rr)}
{R_\rv \, P_0 - \varepsilon \, R_\rd \, P_\rs(T_\rr)} \: .
\nonumber
\end{equation}

\vspace*{-3mm} \noindent
By substituting $q_\rt = q_\rv + q_\rl$, Eq.~(\ref{eq:ssum}) can be written as
\vspace*{-5mm}
\begin{align}
s - s_\ell(T_\rr) & = 
\cpell \: 
\ln \!\left( \frac{T}{T_\rr} \right) 
- R_\rd \: (1-q_\rt) \, 
\ln \!\left( \frac{P_\rd}{P_{\rd,0}} \right) 
\nonumber \\
& - R_\rv \: q_\rt 
\ln \!\left( \frac{P_\rv}{P_{\rv,0}} \right) 
- q_\rl \: (s_\rv - s_\rl).
\label{eq:ref-s-thetal}
\end{align}

\vspace*{-2.5mm} \noindent
In equilibrium $q_\rl \: (s_\rv - s_\rl) = q_\rl \: \ell_\rv/T$, namely with either $q_\rl=0$ or $P_\rv=P_\rs.$ Recasting the above in the form of Eq.~(\ref{eq:sx}), and defining $\theta_\ell$ as the value of the reference temperature that satisfies $s_\ell(\theta_\ell) = s,$ yields:
\vspace*{-2mm}
\begin{equation}
\theta_\ell 
\; = \; \vartheta_{\rx=\ell}
\; = \; 
T \: 
\left(\frac{P_0}{P} \right)^{\kappa_\ell} 
\exp \!\left(-\frac{q_\rl \: \ell_\rv}{\cpell \: T}\right) 
\; \: \Omega_\ell
\: ,
\label{eq:thetal}
\end{equation}
where
\begin{equation}
\Omega_\ell 
\: = \: 
\left(\frac{R}{R_\ell} \right)^{\kappa_\ell} 
\left( \frac{q_\rt}{q_\rv} \right)^{q_\rt R_\rv/\cpell}
\end{equation}
and $\kappa_\ell = R_\ell / \cpell.$ This equation differs slightly from the one derived by Betts (1973) as his Eq.~(8) makes the tacit approximation that $\rd \ln P_\rv = \rd \ln P,$ which neglects contributions from changes in $q_\rv$ for saturated perturbations. Accounting for these effects gives rise to the small correction represented by $\Omega_\ell,$ which departs from unity only when condensate is present in equilibrium. Eq.~(\ref{eq:thetal}) differs fundamentally from \cite{HaufHoeller1987}, who by limiting themselves the $\re$-state composition for the reference state, proposed a $\theta_\ell$-like quantity as an approximation to $\theta_\re.$ 

Physically, $\theta_\ell$ measures the temperature the air would have were any (here liquid) condensate evaporated through a process of isentropic warming by compression. To the extent its reference state is in equilibrium, it is thus a potential temperature in the same (strict) sense as $\theta,$ provided that all condensed water can be evaporated in this reference state.
\vspace*{-2mm}

\subsection{The entropy potential temperature, $\theta_s$} 
\label{sec:thetas}
\vspace*{-2mm}

Despite frequent statements to the contrary, neither $\theta_\ell$ nor $\theta_\re$ are indicative of the specific entropy, $s,$ of moist-air. To address this shortcoming, \cite{Marquet2011QJ} introduced the entropy potential temperature $\theta_s$. The insight required to ensure that $\theta_s$ measures entropy, is the necessity to completely standardize the reference state composition, denoted by $\rx$. For a multi-component system, doing so introduces a dependency on the absolute entropies, and a role for the third law in atmospheric physics (Appendix~B).

To adapt the various derivations of $\theta_s $ to our purposes we begin with the same form of the entropy equation as was used to derive $\theta_\ell,$ namely Eq.~(\ref{eq:ref-s-thetal}), which corresponds also to Eq.~B.10 (with $q_\ri=0)$ of \cite{Marquet2011QJ}. Adopting a reference state for a dry atmosphere, at standard pressure, and defining $\theta_s$ as the value of the reference temperature that satisfies $s_d(\theta_s,P_0) = s$ implies that:
\vspace*{-2mm}
\begin{equation}
\theta_s \; = \; 
\vartheta_{\rx=s} \; = \; 
 \theta \;\:
 \exp\!\left( -\frac{ q_\rl \: \ell_ \rv}{c_{p\rd} \: T}
 + 
 \Lambda \: q_\rt
\right) \; \:\Omega_s \: , 
\label{eq:thetas}
\end{equation}

\vspace*{-4mm} \noindent
where
\begin{equation}
\Omega_s \; = \; 
\left(\frac{R}{R_\re}\right)^{\kappa_\rd} \!
\left(\frac{P_{\rd}}{P_{\rd,0}}\right)^{q_\rt \, \kappa_\rd} \!
\left(\frac{P_{\rv,0}}{P_{\rv}}\right)^{q_\rt \,\gamma}
\!\left(\frac{T}{T_0}\right)^{q_\rt \, \lambda}
\! , \!
\label{eq:Omegas}
\end{equation}
$\gamma = R_\rv / \cpd$,
$\lambda = \cpv/\cpd-1$
and
\begin{equation}
 \Lambda \; = \;
 \frac{ s_\rv(T_0,P_{\rs,0}) - s_\rd(T_0,P_{\rd,0})}{\cpd}
\; \approx \; 5.867 \,. \label{eq:Lambda}
\end{equation}
Reference entropies for water vapor, $s_\rv(T_0,P_{\rs,0}) \approx \SI{12672}{\joule\per\kilo\gram\per\kelvin},$ and dry air, $s_\rd(T_0,P_{\rd,0}) \approx \SI{6778}{\joule\per\kilo\gram\per\kelvin},$ are known up to $\SI{\pm1.5}{\joule\per\kilo\gram\per\kelvin}.$ These uncertainties arise from uncertainties in the standard values (Table~\ref{tab:constants}) from which they are computed. The accuracy of $\Lambda$ is thus of about $\pm 0.003$ unit.
\vspace*{-1mm} \noindent

\begin{table}[t]
 \caption{\it 
Thermodynamic constants calculated with dry air composed with a \ch{CO2} concentration of \SI{420}{ppmv}}
 \label{tab:constants}
\vspace*{-2mm}
 \begin{center}
 \begin{tabular}{lrl}
 \hline \hline
 Constant & Value & Units \\ \midrule
 $T_{0}$ & 273.15 & \si{\kelvin}\\
 $P_{0}$ & 100000 & \si{\pascal}\\
 $P_{\rs,0} = P_\rs (T_0)$ & 611.21 & \si{\pascal}\\
 $\cpd$ & 1004.66 & \si{\joule\per\kilo\gram\per\kelvin}\\
 $\cpv$ & 1865.01 & \si{\joule\per\kilo\gram\per\kelvin} \\
 $\cpl$ & 4179.57 & \si{\joule\per\kilo\gram\per\kelvin} \\
 $R_\rd$ & 287.04 & \si{\joule\per\kilo\gram\per\kelvin} \\
 $R_\rv$ & 461.52 & \si{\joule\per\kilo\gram\per\kelvin} \\
 $s_{\rd,0} = s_\rd(T_0,P_0)$ & 6776.2 & \si{\joule\per\kilo\gram\per\kelvin} \\
 $s_{\rv,0} = s_\rv (T_0,P_0)$ & 10319.7 & \si{\joule\per\kilo\gram\per\kelvin} \\
 $s_{\rl,0} = s_\rl (T_0)$ & 3516.7 & \si{\joule\per\kilo\gram\per\kelvin} \\
 $\ell_{\rv,0} = \ell_\rv (T_0)$ & 2500.93E3 & \si{\joule\per\kilo\gram}\\
\hline
 \end{tabular}
 \end{center}
\end{table}

Physically, $\theta_s$ is the temperature that dry air must have to have the same entropy as the moist system at standard pressure. 
Like $\theta_\re$ and $\theta_\ell$, $\theta_s$ shares the property of reducing to $\theta$ for $q_\rt = 0,$ however in the form of Eq.~(\ref{eq:thetas}) it does so more transparently. This can also be understood due to the relationship
\begin{equation}
s = s_{\rd,0} + \cpd \:\ln\left( \frac{\theta_s}{T_0} \right) \: ,
\label{eq:S_thetas}
\end{equation}
which generalizes the dry-air formula derived by \cite{Bauer1908} to moist-air.
\vspace*{-2mm}

\subsection{Reference states and pseudo-entropies.} 
\vspace*{-2mm}

A substantial and enduring body of literature \citep{PauluisEtAl2008,Pauluis_Czaja_Korty_2010,Raymond2013,Romps2015}, which dates back to \cite{Emanuel1994}, introduces the moist potential temperatures, $\theta_\re$ and $\theta_\ell,$ as a measure of the entropy that would arise if the reference entropies in Eqs~(\ref{eq:entropy-equiv}) and (\ref{eq:ref-s-thetal}) -- respectively depending on $s_\re(T_\rr)$ and $s_\ell(T_\rr)$ -- were assumed to be zero. By adopting this approach one can arrive at expressions for $\theta_\re$ and $\theta_\ell$ that are equivalent to Eqs (\ref{eq:thetae}) and (\ref{eq:thetal}), with the seemingly attractive property that $\theta_\re \propto T_\rr \exp(s/c_\re)$, and equivalently $\theta_\ell \propto T_\rr \exp(s/c_\ell).$ This has led many authors to conclude that $\theta_\re$ and $\theta_\ell$ measures the entropy, or at least a closely related quantity which \cite{Pauluis2018} calls the `relative' entropy.\footnote{As pointed out by \cite{MarquetThibaut2018JAS} this terminology risks confusion with the paper where the \cite{Shannon_1948} entropy is defined, but with another different quantity with the same name of `relative' entropy.} 

A difficulty with defining the moist air entropy as a `relative' entropy, in the sense of \cite{Pauluis2018}, is that it is then measured relative to a reference state that varies with the composition of the system, so that comparing `relative' entropies of fluid parcels invariably conflates differences in their reference state entropies. In a single component fluid, where the composition is fixed, this problem vanishes. To finesse this difficulty, some of the above cited studies have asserted that $s_{\rv,0} - s_{\rd,0},$ which defines $ \Lambda \, \cpd$ in Eq.~(\ref{eq:Lambda}), or analogously $s_{\rl,0} - s_{\rd,0}$ as determines $s_\re(T_\rr),$ can be set arbitrarily without breaking the more general links between $\theta_\ell$, $\theta_e$ and the moist-air entropy. For moist-air, the matter-change entropies differentiate how different forms of matter contribute to the entropy, analagously to the phase-change entropies for different phases of matter. While this is important for specifying the entropy, the absence of conversions (chemical reactions) between the different forms of matter (water and dry air) makes the dynamic role of these matter change entropies less obvious than for the phase change entropies. This doesn't, however, ameliorate the difficulty, which is that if $\theta_\re$ and $\theta_\ell$ purport to measure the entropy relative to something, that \textit{something} needs to be meaningfully specified.


To circumvent these difficulties \cite{Marquet2011QJ} derived $\theta_s.$ In terms of the present interpretative framework, $\theta_s$ can be understood as the result of an open-process that transforms the moist-air to a dry-air reference state by removing the water while heating to maintain constant entropy -- isentropic desiccation. This then defines $\theta_s$ in terms of a reference state whose composition can be fixed absolutely ($q_{\rr,\rt} = 0$), thereby fixing $s_{r,x}$ and $c_\rx$ independently of the state of the parcel, and recovering the desired property whereby $\theta_s \propto T_{\rr,s} \exp(s/c_s),$ as described by Eq.~(\ref{eq:S_thetas}).
\vspace*{-2mm}


 \subsection{Simplified expressions} 
\label{sec:simplified}
\vspace*{-2mm}

The moist potential temperatures are often approximated by neglecting the minor effects of water, i.e., on the thermodynamic constants, or the contribution of the partial vapor pressure to the total pressure. With this approximation $\Omega_\rx \rightarrow 1$ and:
\begin{align}
\tilde{\theta}_\re & \: = \; 
\theta \:\:
\exp\!\left( \frac{+q_\rv \: \ell_\rv}{\cpd \: T}\right) \label{eq:a1} \: , \\
\tilde{\theta}_\ell & \: = \;  
\theta \:\:
\exp\!\left( \frac{-q_\rl \: \ell_\rv}{\cpd \: T}\right) \label{eq:a2} \: , \\
\tilde{\theta}_s & \: = \;  
\tilde{\theta}_\ell \:\: 
\exp\!\left( \, \Lambda \: q_\rt \, \right), \label{eq:a3}
\end{align}
with tilde denoting the approximation. Errors introduced by the use of Eqs.~(\ref{eq:a1})-(\ref{eq:a3}) in lieu of Eqs.~(\ref{eq:thetae}), (\ref{eq:thetal}) and (\ref{eq:thetas}) are on the order of \SI{1}{\percent}, and tend to increase with differences in the pressure from the reference state (see 
Fig.~A1 in the Appendix~A). 

A sense of the errors associated with these approximations is given by Fig.~\ref{fig:dropsondes}, which also plots the approximations (dashed) alongside the actual values for the \EURECA soundings.

Eqs.~(\ref{eq:a1})-(\ref{eq:a3}) are informative as to the differences in the magnitude of the different forms of $\theta_\rx.$ 
Taking $\ell_\rv/(\cpd T) \approx 9$ and $\Lambda \approx 6,$ yields
\begin{equation}
 \ln\left(\frac{\theta_\ell}{\theta}\right) \; \approx \; \ln\left(\frac{\theta_s}{\theta}\right) - 6 \; q_\rt \; \approx \; \ln\left(\frac{\theta_\re}{\theta}\right) - 9 \; q_\rt. \label{eq:magnitudes}
\end{equation}
Because $q_\rt$ is positive definitive, this positions $\theta_s$ at roughly the two thirds position between $\theta_\ell$ and $\theta_\re$, as observed in Fig.~\ref{fig:dropsondes}. 

The different magnitudes of the moist potential temperatures reflect the different degree to which temperature has to compensate differences in the composition of the chosen reference state to maintain the same entropy. Comparing $\theta_e$ to $\theta_\ell$ for instance, shows that a system with all its water in the liquid phase must be much warmer, than the same system with all its water in the vapor phase, if it is to have the same entropy.

Keeping in mind that each of the moist potential temperatures describe the same system, with the same entropy, Eq.~(\ref{eq:magnitudes}) shows how, due to $q_t,$ none of the moist potential temperatures are proportional to one another. And although each describes (approximately) a system with the same entropy, at most one can actually be proportional to entropy, which is a state function whose difference between two points (i.e., states) takes a unique value.
\vspace*{-2mm}

\section{Properties of the moist potential temperatures} 
\label{sec:properties}
\vspace*{-2mm}

By virtue of their derivation, the moist potential temperatures, $\theta_\ell,$ $\theta_s$, and $\theta_\re$ are all potential temperatures in the weak sense of the term, i.e., being the temperature of a reference system with the same entropy as the actual system. Only $\theta_\ell$ qualifies as a potential temperature in the strict sense of the term, i.e., corresponding to a reference temperature accessible by an isentropic and closed transformation of a system in equilibrium.

For $\theta_s$ the reference state has a different composition and thus cannot be attained by a closed system. For $\theta_\re$ the reference state is in mechanical (phase)
disequilibrium. Even for $\theta_\ell,$ mechanical equilibrium of the reference state is only guaranteed for under- or just-saturated water vapor pressure at $T_\rr = \theta_\ell,$ which corresponds to $q_\rt < q_\rs(P_0,\theta_\ell),$ a restriction that is satisfied for most atmospheric conditions. Whereas the composition of the reference state with respect to which $\theta_s$ is defined is independent of the composition of the system it measures, the reference state for both $\theta_\ell$ and $\theta_\re$ is set equal to the composition of the system whose state they measure. 

When comparing how differences in properties are measured by differences in the potential temperatures, it simplifies notation to introduce the difference operator, $\Delta$ defined such that 
\vspace*{-2mm}
\begin{equation}
\Delta \left[ \, f(\chi) \,\right] 
\; \equiv \;
f(\chi_2) - f(\chi_1) \label{eq:delta}
\end{equation} 
\vspace*{-8mm}

\noindent
for two states (enumerated by $1$ and $2$) of any variable $\chi$ and for any function $f.$ Hence $\Delta \left[ \, f(\chi) \,\right] = 0$ if $\chi_2 = \chi_1.$ What distinguishes $\theta_s$ from $\theta_\ell$ and $\theta_\re$ is that $q_{\rr,s}$ is absolute, i.e., it is the same (zero in our case) for all states. Hence $\Delta (q_{\rr,s}) = 0$ by definition. 
\vspace*{-2mm}

\subsection{Entropy} 
\label{sec:entropy}
\vspace*{-2mm}

For moist-air and from Eq.~(\ref{eq:sx}), denoting by $\theta_\rx$ the temperature for the reference state whose composition is chosen by $\rx$ such that $s = s_\rx(\theta_\rx),$ it follows that 
\vspace*{-2mm}
\begin{equation}
 \Delta s = \Delta [s_\rx(\theta_x)] \: .
\end{equation}

\vspace*{-2.5mm} \noindent
For the case of dry air, the reference entropy (as measured relative to standardized values) with the temperature chosen in this way, satisfies
\vspace*{-2mm}
\begin{equation}
\hspace*{-4mm}
s_{\rd,\rr} = 
s_\rd(\theta,P_\rr) = 
s_{\rd,0} 
\,+\, \cpd \ln\!\left(\frac{\theta}{T_0}\right) - R_\rd \ln\!\left(\frac{P_\rr}{P_0}\right)
 \! . \!
\label{eq:entropy-dry}
\end{equation}

\vspace*{-2.5mm} \noindent
Hence the difference between two reference states is given by
\vspace*{-2mm}
\begin{equation}
\Delta s_{\rd,\rr} \; = \; 
c_{p_\rd} \: \Delta \left[\, \ln(\theta) \,\right] 
\: - \: 
R_\rd \: \Delta \left[\, \ln(P_r) \,\right] \: . 
\label{eq:entropy-gas}
\end{equation}

\vspace*{-2.5mm} \noindent
These equations show that by standardizing the reference pressures (so that $P_{\rr,1}= P_{\rr,2}$), $\Delta s \propto \Delta (\ln \theta),$ hence differences in the dry-air potential temperature measure dry-air entropy differences. 

Let us show that the same is not true for $\theta_\ell$ and $\theta_\re.$ Consider first the entropy differences between two liquid-less ($q_\rt=q_\rv$) reference states (per the definition of the $\ell$-state, both are defined relative to standardized pressures, but can differ in composition, i.e., $q_{\rt,1} \ne q_{\rt,2}$). In this case, after a little bit of algebra, one can show that
\vspace*{-2mm}
\begin{equation}
\Delta [s_\ell (\theta_\ell)] = 
 c_{\ell,2} \:
 \Delta \left[\: \ln(\theta_\ell) \:\right] 
 + \Phi_\ell(q_{\rt,2},q_{\rt,1}) 
 \label{eq:entropy-ref-moist}
 \: ,
\end{equation}
\vspace*{-8mm}

\noindent
where,
\vspace*{-3mm}
\[
\begin{split}
\Phi_\ell & (q_{\rt,2}, q_{\rt,1}) = 
 \left[ \,
 \cpd \,\Lambda
 + (\cpv-\cpd) \ln \!
 \left(\frac{\theta_{\ell,1}}{T_0}\right) 
 \right] \Delta q_\rt \\
& + R_\rd \,\Delta \! \left[\,
 (1+\varepsilon \,q_\rt) \ln (1+\varepsilon \,q_\rt) 
 - (1-q_\rt) \ln(1-q_\rt) 
 \,\right] \\
& + R_\rv \,\Delta \!\left[\,
 q_\rt\ln q_\rt - q_\rt\ln \!\left(\varepsilon + 1\right) 
 \,\right].
\end{split}
\]
\vspace*{-4mm}

Eq.~(\ref{eq:entropy-ref-moist}) shows that, even after standardizing the reference-state pressures the relationship between $\Delta s$ and $\Delta (\ln \theta_\ell)$ is modified by $\Phi_\ell,$ whose value depends on differences in the composition of the two states.
Because $\Phi_\ell(q_{\rt,2}, q_{\rt,1})=0$ only if $q_{\rt,2} = q_{\rt,1},$ 
differences in $\theta_\ell$ can only measures the entropy differences of systems with the same composition ($q_{\rt,2} = q_{\rt,1}$). 
The same is true for $\theta_\re$ although the form of $\Phi_\re$ differs from that of $\Phi_\ell.$

In contrast, for $\theta_s,$ by virtue of adopting a dry reference state (and standardizing the pressures),
\begin{equation}
\Delta[s_s(\theta_s)] = 
 c_{p_\rd} \: \Delta \left[ \, \ln(\theta_s) \, \right] \: .
\end{equation}
To attain this property it is necessary to standardize the reference state composition, i.e., to pick a reference state composition $\rx$ that is fixed independently of the composition of the given state. This of course means that isentropic transformations to this state cannot be closed. So while it is possible to define a moist potential temperature (in the weak sense) that measures entropy ($\theta_s$), no moist potential temperature defined in the strict sense of the term can do so. 
\vspace*{-2mm}

\subsection{Enthalpy} 
\vspace*{-2mm}

Here we investigate to what extent the choice of a moist potential temperature influences its ability to measure enthalpy differences. For the dry potential temperature, $T=\theta$ at the reference state pressure, and so differences in $\theta$ measure differences in the reference state enthalpies, i.e., 
\begin{equation}
\Delta h = c_{p_\rd} \, \Delta \theta . 
\label{eq:warmegehalt}
\end{equation}
It was in this sense that Helmholtz identified $\theta$ with what he called the W\"armegehalt (heat content). 

Similar to the case for entropy, for moist-air, compositional differences influence enthalpy differences in ways that the moist potential temperatures do not fully account for. For moist-air, 
\begin{align}
\!\!
h & = c_\ell \, (T-T_0) 
 - \ell_v \, q_\rl
 + h_{\rd,0} 
 + ( h_{\rv,0}-h_{\rd,0}) \, q_\rt 
 \label{eq:hl} \\
\!\!
 & = c_\re \, (T-T_0) 
 + \ell_v \, q_\rv
 + h_{\rd,0} 
 + ( h_{\rl,0}-h_{\rd,0}) \, q_\rt ,  \label{eq:he}
\end{align}
where a discussion of the reference enthalpies are given in \cite{Marquet2017JAS}. In atmospheric studies the `relative' enthalpies 
\begin{equation}
h_\ell \; = \; 
c_\ell \: T \: - \: \ell_\rv \: q_\rl 
\quad \mbox{and} \quad
h_\re \; = \; 
c_\re \: T \: + \: \ell_\rv \: q_\rv \: ,
\label{eq_hl_he}
\end{equation}
which form parts of Eq.~(\ref{eq:hl}) and Eq.~(\ref{eq:he}), are often introduced, and serve as the enthalpic contributions to the liquid water and moist static energies respectively.

Given that $\theta_\ell$ and $\theta_\re$ respectively measure the temperature, $T,$ at condensate-free and vapor-free reference states, it follows that changes in the reference state enthalpy (not to be confused with the reference quantities, which are denoted by subscript 0) can be written as
\vspace*{-2mm}
\begin{align}
\!\! \Delta (h) 
 & \: = \; \Delta (\cpell \, \theta_\ell) 
 \: + \: L_{\ell,0} \: \Delta(q_\rt) 
 \label{eq:dhl} \\ 
 & \: = \; \Delta (c_\re \, \theta_\re) 
 \: + \: L_{\re,0} \: \Delta(q_\rt), 
 \label{eq:dhe} 
\end{align}
with $L_{\ell,0}$ and $L_{\re,0}$ constant reference quantities readily deduced from Eqs.~(\ref{eq:hl}) and (\ref{eq:he}).

Eqs.~(\ref{eq:dhl}) and (\ref{eq:dhe}) demonstrate how the dependence of the reference state enthalpy on $q_\rt$ conflates the relationship between $\Delta (h)$ in the reference state and $\Delta (\cpe \theta_\re).$ The situation for $\theta_\ell$ is no different. However, for many purposes (e.g., measuring temperature changes from mixing) differences in reference temperatures and enthalpies play no role -- knowledge of differences in `relative' enthalpies and $q_\rt$ is sufficient. From Eqs.~(\ref{eq_hl_he}) we note that in the reference state $ \Delta(h_\re) \rightarrow \Delta (\cpe \, \theta_\re)$ and $\Delta(h_\re) \rightarrow \Delta (c_\ell \, \theta_\ell)$. This gives a weak form of correspondence between $\theta_\re$ or $\theta_\ell$ and their dry air counterpart, $\theta.$ 

Because $\theta_s$ defines the temperature dry air must have to have the same entropy as the moist system, differences in $\Delta \theta_s$ measure differences in the enthalpy of dry air with the same entropy as the moist systems being compared, but the meaning of this enthalpy is not especially informative. This is is not unexpected given that $\Delta(\theta_s)$ was designed to measure changes in entropy, not enthalpy.
\vspace*{-2mm}

\subsection{Linear mixing} 
\vspace*{-2mm}

Entropy $S$ and enthalpy $H$ for a given mass ($m$) are both extensive variables, whereas specific values for both entropy ($s=S/m$) and enthalpy ($h=H/m$) are intensive variables. The total entropy and enthalpy embodied in two parcels of air of mass $m_1$ ad $m_2$ is the sum of the entropy and enthalpy of each parcel, respectively. When the parcels mix, the total entropy increases because the process is irreversible, but the total enthalpy doesn't change. Therefore the specific enthalpy is linearly mixing, but the specific entropy is not.

Therefore, in addition to labelling entropy, being conserved along isentropes and measuring enthalpy differences at constant pressure, $\theta \: = (p_0/p)^\kappa \: T$ also mixes linearly for dry air at constant pressure. By this it is meant that if one lets $m_1$ and $m_2$ denote the specific masses of two air-parcels, whose states are indicated by the enumeration ($1$ or $2$), then upon mixing masses of dry-air the value of $\theta$ is just the mass-fraction weighted sum of the constituents, i.e., 
\vspace*{-1mm}
\begin{equation}
\theta = \theta_1 + \eta \: \Delta (\theta)
\quad \text{where} \quad
 \eta = \frac{m_2}{m_1+m_2}.
\label{eq:thetax1}
\end{equation}
This property of linear mixing is desirable of quantities used in numerical models. It is verified for the mixing of both enthalpy, and `relative' enthalpy, for dry and moist air, and extends to $\theta$ for the case of dry air. For this property to also be transferable to the $\theta_\rx$ thus requires $\Delta h_\rx = c_\rx \,  \theta_\rx.$ 
\vspace*{-2mm}

From the discussion of the previous section, this would seem to be the case for $\rx \in \{\re,\ell\}.$ However, as pointed out there, $\Delta (h_\rx) \rightarrow c_\rx \, \theta_\rx$ only for the reference state. For mixing of air in a different state it is additionally required that the work done to move the mixed system from its reference state to the given state is the same as the work done on the component systems to move them to their reference state, that this is not generally satisifed is also why the moist static energies do not mix linearly \citep{Bretherton1987}. 
\vspace*{-2mm}

\section{Examples} \label{sec:Examples} 
\vspace*{-2mm}

In this section we present several examples chosen to further illustrate the properties of various choices of $\theta_\rx$. The first compares the structure of the tropical atmosphere as seen through profiles of $\theta_\re,$ $\theta_{\ell}$ and $\theta_s.$ The second explores the ability of $\theta_\rx$ to measure changes in the state of the atmosphere resulting from the isobaric mixing of air-parcels, using a challenging but relevant example of cloud top mixing. The third compares the ASTEX observed vertical profiles of $\theta_\re,$ $\theta_{\ell}$ and $\theta_s$ to study the transition from stratocumulus to cumulus.
\vspace*{-2mm}

 \subsection{Contrasting the wet and dry tropics} 
 \label{--Subsection:VerticalProfiles}
 \vspace*{-2mm}

For the first example we compare the representation of the thermodynamic state in the troposphere in terms of $\theta_\re,$ $\theta_\ell$ and $\theta_s.$ Composite temperature and humidity profiles are derived from global storm-resolving (\SI{2.5}{\kilo\meter}) simulations from the DYAMOND project \citep{StevensEtAl2019c} using the ICON model \citep{HoheneggerEtAl2020}. The composite soundings are taken points over the ocean within the deep (\ang{10}S-\ang{10}N) tropics. Two soundings are constructed, the first by compositing over regions drier than the 10th percentile of precipitable water, the second by compositing over columns moister than the 99th percentile of precipitable water (to capture the very moistest convective regions). They thereby contrast the thermodynamic structure of the dry and wet tropics, the latter being indicative of regions of active convection.

\begin{figure}[htb]
\includegraphics[width=0.99\linewidth]{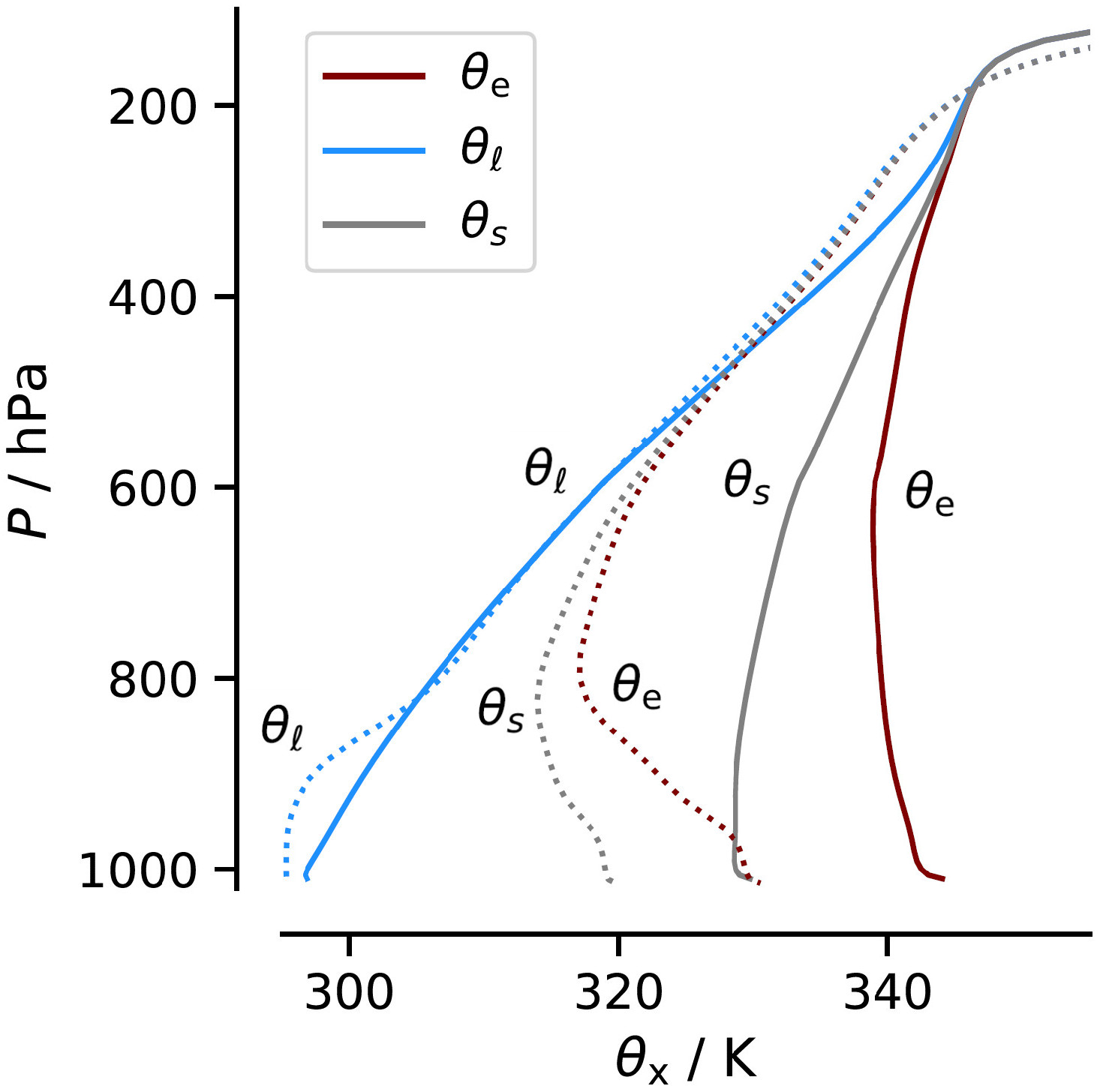}
\vspace*{-2mm}
\caption{\it \small 
Profiles of $\theta_\re,$ $\theta_\rl$ and $\theta_s$ for a composite sounding over the dry (dotted) and moist (solid) tropics.} 
\label{fig:profiles}
\end{figure} 

Fig.~\ref{fig:profiles} complements Fig.~\ref{fig:dropsondes} to more generally show how different expressions for $\theta_\rx$ have neither the same values, nor even the same structure. If each expression for $\theta_\rx$ were proportional to the entropy (or the entropy as measured relative to some reference), as is sometimes maintained, then how in the case of the moist atmosphere (solid lines) could $\theta_s$ increase in the lower atmosphere (between \SIrange{800}{600}{\hecto\pascal}) while $\theta_\re$ decreases. Likewise, how can $\theta_s$ decrease below \SI{800}{\hecto\pascal} in the dry sounding (left panel) where $\theta_\ell$ increases. This provides a vivid example of how differences in $\theta_\re$ and $\theta_\ell,$ measure differences in the entropy of the reference states of each profile, rather than differences in the entropy of the actual state. Put another way, if two gas quanta have the same entropy, but differ in composition, their values of $\theta_\re$ and $\theta_\ell$ will vary to reflect these differences in composition.

\begin{figure*}[htb]
\centering\noindent
\includegraphics[width=0.99\linewidth]{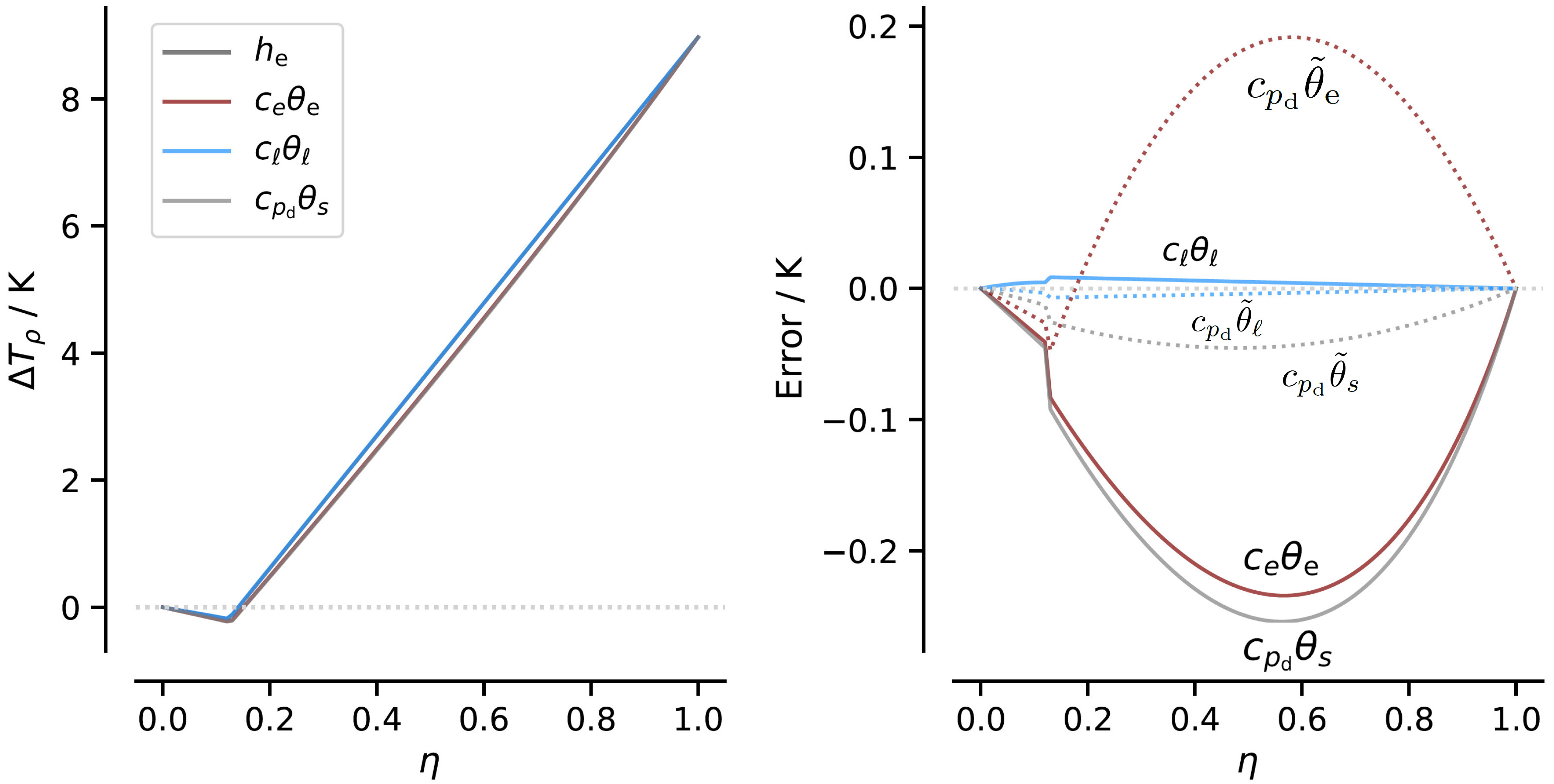} 
\vspace*{-4mm}
\caption{\it \small 
Buoyancy perturbation between isobaric mixing of saturated ($T=\SI{283.02}{\kelvin}$ and $q_\rt = \SI{9.0}{\gram\per\kilo\gram}$) and unsaturated air ($T=\SI{292.94}{\kelvin}$ and $q_\rt = \SI{1.5}{\gram\per\kilo\gram}$) at \SI{920}{\hecto\pascal} with a cloud-top water content of \SI{0.7}{\gram\per\kilo\gram}. The buoyancy temperature derived by mixing parcels based on their moist `relative enthalpy', $h_\re,$ is plotted versus mixing fraction, along with the error. Errors (right panel) are compared for the mixing of the exact $c_\rx \theta_\rx$ (solid) versus the simplified $\cpd \tilde{\theta}_\rx$ (dotted) versions. Color coding is indicated in the key in the left panel.}
\label{fig:mixing2}
\end{figure*}

In contrast, by virtue of being defined relative to an absolute reference state, $\theta_s$ is proportional to $\exp(s / \cpd).$ For both the dry and the moist soundings, profiles of $\theta_s$ vary less with height than profiles of $\theta_\ell$ or $\theta_\re.$ The inference being that the entropy, $s,$ is better mixed in the lower troposphere than one would surmise by associating it with $\theta_\re$ or $\theta_\ell.$ And whereas in the dry atmosphere a surface thermal source and an above-PBL radiative sink of entropy are associated with an entropy minimum near (\SI{800}{\hecto\pascal}), the entropy is everywhere increasing in the convective sounding. The profile of $\theta_s$ in particular emphasizes that the entropy of the lower troposphere is relatively constant, but that the dry regions have an entropy deficit as compared to the moist regions, presumably due to the radiant energy sink as air slowly subsides away from regions of active convection. These properties are only possible to ascertain from the profile of $\theta_s.$ 

As evident from Eq.~(\ref{eq:magnitudes}) $\theta_\re$ is consistent with constant $\theta_s$ only for the case of constant $q_\rt.$ Homogenizing $\theta_\re$ while reducing $q_\rt,$ as the moist profiles in Fig.~\ref{fig:profiles} show to be the case in the convective state, increases $\theta_s.$
\vspace*{-2mm}

 \subsection{Cloud-edge isobaric mixing} 
\label{--Subsection:IsobaricMixing}
\vspace*{-2mm}

For the second example we compare isobaric mixing between two air-masses at a cloud-top interface. The mixing of saturated and unsaturated air is non-linear, so this provides a challenging but relevant test of the properties of the various forms of $\theta_\rx$. The case we explore is based on measurements of marine stratocumulus made as part of the DYCOMS-II field study, wherein a stratocumulus layer was topped by warmer and much drier air \citep{StevensEtAl2003}. The conditions sampled during the first research flight satisfied the buoyancy reversal criteria, whereby the air aloft, which we designated by subscript 1, had a higher density temperature, $T_\rho$ than the air in the cloud, designated by subscript 2. This situation, whereby $T_{\rho,1} > T_{\rho,2}$ corresponds to a stable stratification in the absence of mixing. For the observed conditions, mixtures of the warmer drier air aloft with the cooler saturated air in the cloud layer, would (for a range of mixing fractions) result in air-parcels denser than the air in the cloud layer. This is a mixing instability whose importance for the dynamics of marine stratocumulus continues to be debated \citep{Deardorff80,Randall80,Mellado2017}.

We calculate $\Delta T_\rho = T_\rho(\eta) - T_{\rho,2}$ as a function of the mixing fraction $\eta.$ As defined as in Eq.~(\ref{eq:thetax1}), $\eta$ denotes the specific mass of one component (which we denote by subscript 2) of a binary mixture. For the reference (black line) we calculate the properties of the mixed air by virtue of both $q_\rt$ and $h_e$ (as given by Eq.~(\ref{eq:he})) mixing linearly at constant pressure $P$, 
leading to
\begin{align}
q_\rt & = q_{\rt,1} + \eta \: \Delta(q_\rt) \, ,
\label{eq_mixing_qt} \\
h_\re & = h_{\re,1} + \eta \: \Delta(h_\re) \, .
\label{eq_mixing_he}
\end{align}
Together with the fixed pressure $P$, this defines the state of the system, from which $T$ and $T_\rho$ can be calculated. 

Fig.~\ref{fig:mixing2} confirms our earlier arguments that none of the formulations for $\theta_\rx$ linearly mix. Although our particular example involves phase changes, the structure of the error in Fig~\ref{fig:mixing2} (right panel), which is on the order of \SIrange{5}{10}{\percent} and maximizes (near $\eta=0.6$) for unsaturated mixtures, is primarily due to the effect of $\Delta q_\rt$ rather than from phase changes.

This analysis serves as a reminder that isentropic invariance (conservation) of a thermodynamic quantity does not guarantee that it mixes linearly. For the case of $\theta_s$ this should be clear, as the mixing itself is a source of entropy and $s \propto \ln(\theta_s).$ Adopting a log mixing rule, i.e. 
\begin{equation}
 \theta_s(\eta) =
 \exp\left\{\,
 \ln(\theta_{s,1})
 + \eta \: 
 \Delta\left[ \,
 \ln(\theta_s)
 \, \right]
 \, \right\} \: ,
 \label{eq:logrule}
\end{equation}
is thus equivalent to (incorrectly) assuming linear mixing of $s.$
This, at least, gives the resultant temperature errors a physical interpretation, i.e., that which arises from neglecting the entropy production through mixing \citep{Richardson_19a}. Errors incurred by linearly mixing $\theta_\re$ or $\theta_\ell$ are more challenging to interpret.

 \subsection{Stratocumulus-Cumulus transition} 
\label{sec:ScCuTransition} 
\vspace*{-2mm}

\begin{figure*}[htb]
\centering\noindent
\includegraphics[width=0.99\linewidth]{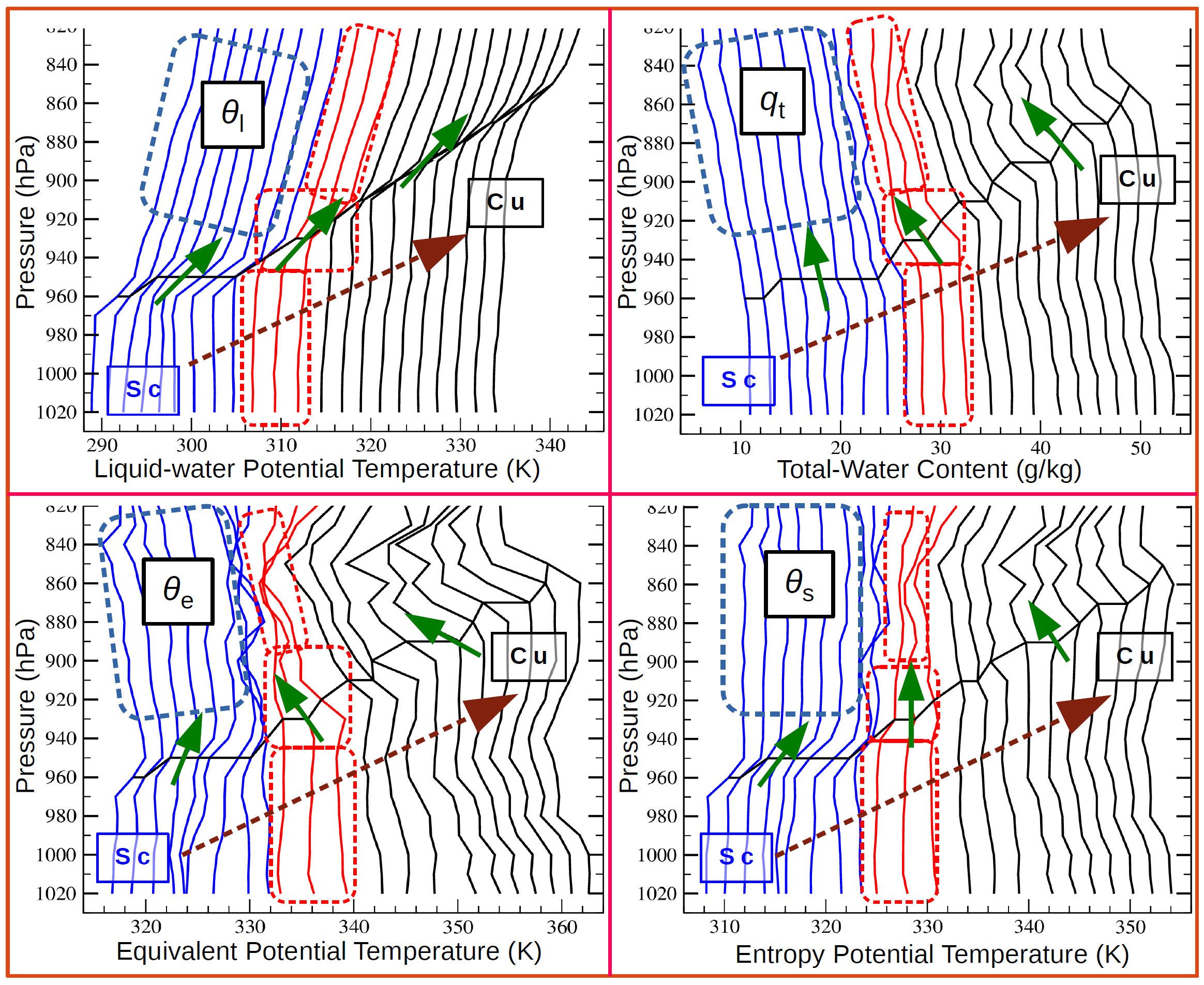}
\caption{\it \small Vertical profiles of $\theta_\re,$ $\theta_{\rl}$ and $\theta_s$ plotted for half of the observed sounding of the first ASTEX Lagrangian experiment (with a shift of \SI{2}{\kelvin} or \SI{2}{\g\per\kg} between each profiles). Stratocumulus (Sc) profiles are colored blue, whereas Cumulus (Cu) profiles are colored black. Transition profiles between the two regimes are colored red, with the purple arrow indicating the deepening of the PBL associated with the transition. The green arrows show the sign of the top-PBL jump for each variable and for each regime: positive if tilted to the right, null if vertical, negative otherwise. The blue and red dashed boxes have been added to highlight the isentropic regions where $\theta_s$ (and not $\theta_e$) is constant despite the opposite vertical gradients in $\theta_l$ and $q_t$ which compensate with the special value of $\Lambda$ given by the third law.}
\label{fig:ASTEX-profiles}
\end{figure*}

For the third example we study profiles of $\theta_\rx$ for the forty-three observed sounding profiles of the first ASTEX Lagrangian experiment described in \cite{BrethertonPincus95} and \cite{deRoodeal97}. The profiles are shown for the respective values of $\theta_\rx$ in Fig.~\ref{fig:ASTEX-profiles}. The sets of profiles are subjectively associated with different cloud regimes. Stratocumulus profiles (colored blue) are associated with mixing from cloud top to the surface and have
extensive cloud cover. Profiles associated cumuliform cloud regimes, are colored black. The transition between the two, often associated with stratocumulus whose thermodynamic properties are differentiated (decoupled) from the thermodynamic state of the sub-cloud layer, are colored red. 

As a consequence of the changing profile of $q_\rt,$ the cloud transition admits very different interpretations depending on which form of $\theta_\rx$ it is viewed from. Transition profiles are associated with a weakening of the negative $\theta_\ell$ gradients in the hydro-lapse\footnote{The term hydro-lapse is used to demarcate the trade-wind inversion region as the fall off of moisture with height is often more pronounced than the increase of temperature at the top of the trade-wind cloud layer.} regions that demarcates the top of the marine (moist) layer, and a reversal above a certain threshold value of the gradient as measured by $\theta_\re.$ The latter is the basis for the cloud-top entrainment instability hypothesis \citep{Randall80}. The behavior of $\theta_s$ is somewhat different, as the transition is better demarcated by a homogenization of $\theta_s$ in the lower troposphere and almost a null top-PBL jump. Whether this is the cause as once suggested by \cite{Richardson_19a}, or an effect, of increased lower tropospheric mixing is difficult to say, particularly given the strong entropy sources and sinks in this region of the atmosphere. Nonetheless the observation, whereby $\theta_s$ gradients tend to vanish as stratocumulus gives way to shallow cumulus, has recently been used by \citet{MarquetBechtold2020} to introduce an index for demarcating regions of stratocumulus from cumulus.
\vspace*{-2mm}

\section{Conclusions} \label{sec:conclusions}
\vspace*{-2mm}

Our main conclusion is that it is hard to avoid accounting for composition when comparing air-parcels whose composition varies. While this might seem trivial, a poor recognition of this fact can, and has, led to considerable confusion -- for instance the idea that somehow $\theta_\re$ measures entropy.

In retrospect it seems obvious that composition matters for varied air-parcel properties in ways that the introduction of a single moist potential temperature cannot account for -- a point also emphasized by \cite{PauluisEtAl2008}. Recognizing this fact raises the question as to whether the different moist potential temperatures measure the same thing, and if not what precisely do they measure? 

\clearpage

We answer these questions first by showing that the equivalent potential temperature ($\theta_{\re}$) of \cite{Rossby1932}, the liquid-water potential temperature ($\theta_\ell$) of \cite{Betts1973} and the entropy potential temperature ($\theta_s$) of \cite{Marquet2011QJ} all share the property of describing the temperature air in some specified reference state would need to have, to have the same entropy as the air-parcel they characterize. Each of these adopt standard pressure for the reference state, but differ in the disposition of the variable component. The reference state for $\theta_s$ is water-free, the reference state for $\theta_\ell$ is condensate free, and the reference state for $\theta_\re$ is vapor-free. 

Even if it is not crucial to the validity of its definition, only the $\theta_\ell$ reference state is attainable through an isentropic, reversible, and closed transformation, as is the case for the dry potential temperature, $\theta,$ and then only in the case when the mass fraction of the water mass in the air-parcel is less than the saturated mass fraction at the reference state temperature and pressure. The reference state for $\theta_\re$ is one of mechanical (phase) disequilibrium of the water phase, and the reference state for $\theta_s$ can only be accessed by an open process (to remove the water mass entirely).

The reference states that define $\theta_\ell$ and $\theta_\re$ are variable, which means they depend on the composition of the parcel which they characterize. In contrast, the reference state of $\theta_s$ is absolute; it is independent of the composition of the air-parcel it characterizes. The latter is a necessary condition for a moist potential temperature to measure entropy. Put differently, $\theta_\re$ (and $\theta_\ell$) only measures entropy differences of air-parcels with the same composition, hence in a variable composition atmosphere, only isopleths of $\theta_s$ coincide with isentropes. Compositional contributions to the entropy are substantial and can only be accounted for by accounting explicitly for the entropy difference between dry air and water vapor (via $\Lambda$), similar to the well appreciated fact that condensational effects can only be accounted for by explicitly accounting for entropy differences between water vapor and condensate, which in equilibrium is proportional to $\ell_\rv,$ the vaporization enthalpy. This is why $\ln(\theta_\re/\theta_\ell) \propto q_\rt \: \ell_\rv$ and why $\ln(\theta_s/\theta_\ell) \propto \Lambda \: q_t,$ with $\Lambda$ measuring the difference between the entropy of water-vapor and dry-air. 

It should come as no surprise that each of the moist potential temperatures are useful for precisely measuring something, and each usefully approximates several air-parcel properties, but none usefully approximate all important properties. $\theta_\ell$ and $\theta_\re$ are poor measures of entropy, but accurately measure the reference state `relative' enthalpy. In the case of $\theta_\re$ whose reference state has already valorized the vaporization enthalpy of its water, the addition or removal of condensate, has a relatively minor effect. Likewise $\theta_\ell$ is relatively insensitive to changes in vapor. This explains the popularity of $\theta_\re$ as a basis for tracking air parcels in the presence of precipitation, or the use of $\theta_\ell$ in studies more interested in isolating an air-mass' thermal properties -- for instance as a component of a mixing diagram. In contrast, $\theta_s$ measures the entropy of moist air. None of the moist potential temperatures mix linearly, and the errors encountered by assuming they do so can be substantial (ranging from a few to ten percent).

Several examples are explored as a basis for exploring trade-offs in the use of different forms of $\theta_\rx$ to interpret the structure of the tropical atmosphere. These examples show how $\theta_s$ is generally better mixed through the tropical troposphere than is either $\theta_\re$ or $\theta_\ell,$ and that the transition from stratocumulus to cumulus is associated with a transition of the troposphere to a state where $\theta_s$ becomes mixed through the lower troposphere, despite considerable gradients in moisture -- whether or not this structure, which is also corroborated by many other observations \cite[see][]{Marquet2011QJ}, is indicative of a process that acts to homogenize entropy, or occurs by chance, is an open question.
\vspace*{-2mm}

 \begin{center} \rule[0mm]{7.cm}{0.1mm} \end{center}
\vspace*{-4mm}

\subsection*{Acknowledgements:} 
\vspace*{-2mm}


The ideas were developed jointly by the authors and hence ordered alphabetically by last name. 
The research was made possible by generous public support for the scientific activities of the 
Max Planck Society, M\'et\'eo-France and the CNRS
-- sometimes research is still possible without third party funding. 
The authors thank Dave Raymond, Martin Singh and an anonymous reviewer, 
as well as the editor 
William Boos, for their constructive and critical comments which led to substantial 
improvements in the presentation of our ideas.

\renewcommand{\thefigure}{A\arabic{figure}}
\setcounter{figure}{0}
\begin{figure*}[htb]
\centering\noindent
\includegraphics[width=0.99\linewidth]{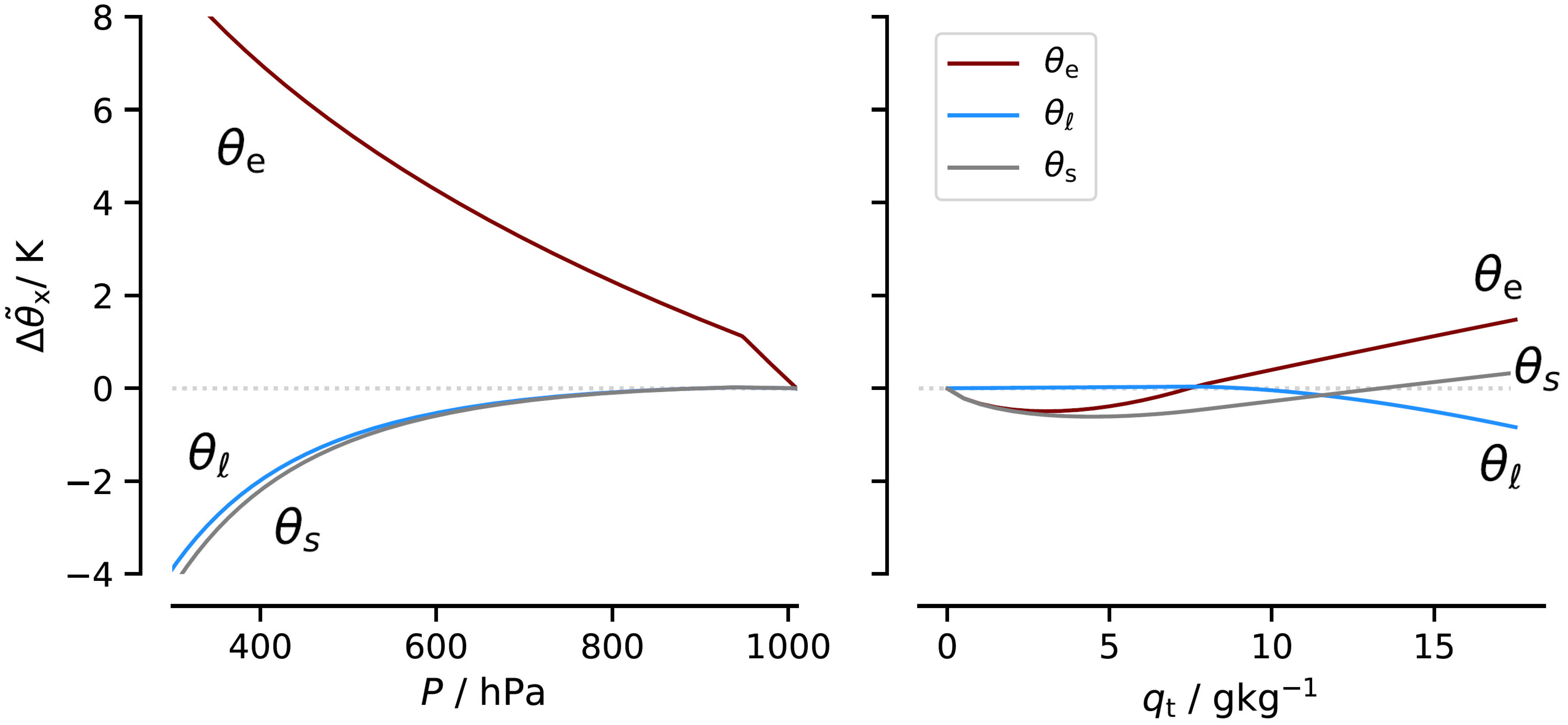}
\vspace*{-4mm}
\caption{\it \small 
Approximation errors associated with expressions of $\tilde{\theta}_\rx.$ Variations of 
$\tilde{\theta}_\rx-\theta_\rx (P=\SI{1010}{\hecto\pascal})$ along a saturated isentrope 
(left); $\tilde{\theta}_\rx-\theta_\rx$ as a function of $q_\rt$ for $(T,P)$ =(\SI{280}{\kelvin}, 
\SI{800}{\hecto\pascal}) (right). Lines in the left panel share the color-key of the right panel.}
\label{fig:approximations}
\end{figure*}

%
\vspace*{2mm}
Profiles used for the cumulus to stratocumulus transition are available at 
\url{http://www.atmos.washington.edu/~breth/astex/lagr/README.hourly.html} 
and \url{ftp://eos.atmos.washington.edu/pub/breth/astex/lagr/lagr1/hourly/}. 
\EURECA data for Fig~1 are available from the JOANNE data set as cited. 
The temperature soundings used Fig.~2 are provided courtesy of the DYAMOND project, 
it and a notebook containing the calculations presented in the manuscript are 
made available by the host authors corresponding institutions by contacting 
publications@mpimet.mpg.de.

%


\vspace*{-2mm}

 \begin{center} \rule[0mm]{7.cm}{0.1mm} \end{center}
\vspace*{-2mm}

\noindent
{\bf{{Appendix~A}. Numerical evaluation}.} \label{sec:approximations}

We tested the numerical implementation of expressions for the different forms of $\theta_\rx$.
To do so we numerically integrated the adiabatic form of the first law for our composite system, 
along the (\SI{300}{\kelvin}, \SI{1010}{\hecto\pascal}, \SI{17}{\gram\per\kilogram}) isentrope, 
which also served as the initial condition. 
The integration was terminated when $P$ reached \SI{150}{\pascal}, at which point 
$T=\SI{207.42}{\kelvin}.$ 
We then calculated $\theta_\re,$ $\theta_\ell$ and $\theta_s$ for different tuples of $(T,P),$ 
along this isentrope. 
If properly constructed, the different versions of $\theta_\rx$ should 
adopt different values, but each should be invariant on this isentrope. 
Variations were verified to be smaller than the tolerance of the ODE solver (lsoda) 
used for the numerical integration of the adiabatic form of the first law. 
To achieve this level of accuracy it was necessary to use a relationship for $P_\rs(T)$ 
consistent with the approximations outlined in \S~\ref{sec:preliminaries}, 
and to ensure the adequacy between the definition of $P_\rs(T)$ and the variations, 
or not, of 
$\ell_v(T) = (\cpv-\cpl)(T-T_0)+\ell_v(T_0)$ 
with temperature, due to
\begin{equation}
\frac{1}{P_\rs(T)} \frac{dP_\rs(T)}{dT}
 \; = \; \frac{\ell_v(T)}{R_v\:T^2} \: .
\nonumber
\end{equation}
Using more exact approximations to $P_\rs,$ which account for variations in the specific heats $\cpv$ and $\cpl$ with temperature, introduces inconsistencies in the form of centi-kelvin discrepancies between the temperatures derived by direct integration, and those implied by constant $\theta_\rx$. 

The approximations given by Eqs.~(\ref{eq:a1})-(\ref{eq:a3}) introduce errors on the order of \SI{1}{\percent}, or about $\SI{4}{\kelvin}.$ The left panel of 
Fig.~A1
shows how $\tilde{\theta}_\rx$ changes from its value at \SI{1010}{\hecto\pascal} as pressure is reduced and $\theta_\rx$ is held constant. Through most of the lower troposphere ($P>\SI{600}{\hecto\pascal}$) both $\tilde{\theta}_\ell$ and $\tilde{\theta}_s$ are approximately constant along the isentrope. Errors in $\tilde{\theta}_\re$ are more severe and systematic with pressure. All forms of $\theta_\rx$ show errors in the upper troposphere, but in this region of the atmosphere strong departures from equilibrium associated with the ice phase likely introduce even larger errors, or at least substantial uncertainty. 

The chosen forms for $\tilde{\theta}_\rx$ neglect humidity effects in all terms other than the ones carrying the dominant sensitivity to humidity. For this reason, in the right panel of 
Fig.~A1
we present an evaluation of the errors in the expressions for $\tilde{\theta}_\rx$ as a function of $q_\rt.$ The errors associated with each approximation are small (\SI{1}{\percent}) and commensurate. Certainly we see no basis for choosing one or the other form of $\tilde{\theta}_\rx$ based on it being a better approximation to the true value.

\renewcommand{\thefigure}{B\arabic{figure}}
\setcounter{figure}{0}
\begin{figure*}[hbt]
\centering
\includegraphics[width=0.7\linewidth]{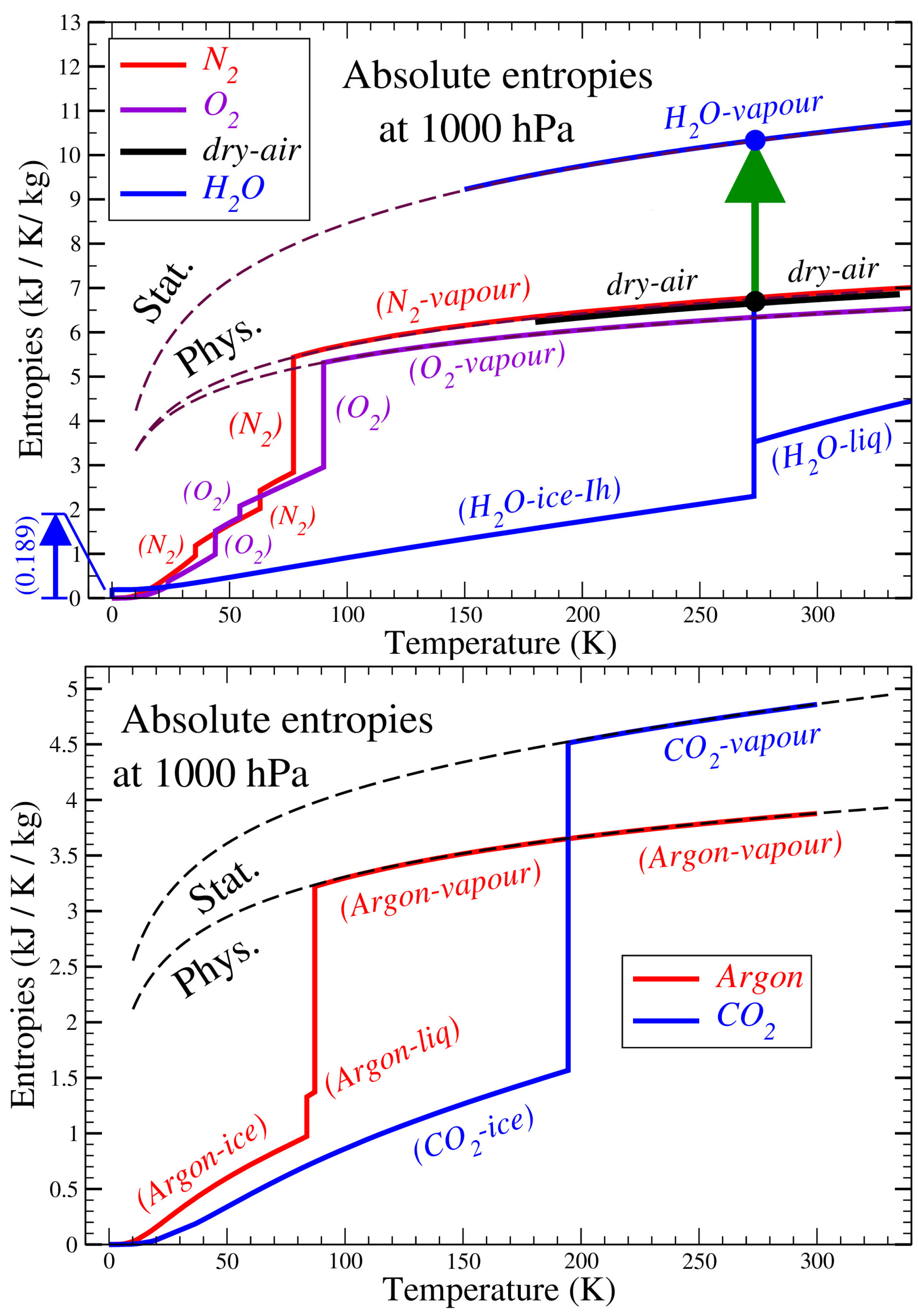}
\caption{\it \small 
Entropies for dry-air (\ch{N2}, \ch{O2}, \ch{Ar}, \ch{CO2} and water \ch{H2O)} species plotted against the absolute temperature and computed at \SI{1000}{\hecto\pascal}. The calorimetric method ($\int_0^T c_p(T') \rd\ln(T') + \sum_j \ell(T_j)/T_j$) corresponds to the coloured solid lines. The third-law hypothesis is applied at $0$~K with zero entropies for all the solid phases, but with the residual entropy of \SI{189}{\joule\per\kilo\gram\per\kelvin} for ice-Ih. The vertical jumps correspond to phase changes at $T_j$ with the phase-change enthalpies $\ell(T_j)$ between solids phases (for \ch{N2} and \ch{O2}), then from solid to liquid phases, then from liquid to vapour phases. The statistical-physics values (black dashed lines) are computed from $S = k\: \ln(W)$ and $F = - \: k \: T \: \ln(Z)$ for the vapour phases according to the method described in \cite{Chase_1998} for translational, rotational, vibrational and electronic partition functions ($Z$).}
\label{figs-entropies}
\end{figure*}

\clearpage

The results shown in this appendix are in agreement with errors on the order of \SI{0.6}{\K} or \SI{0.2}{\percent} shown in Fig.~\ref{fig:dropsondes}. The larger errors in 
Fig.~A1
are due to cumulative effects during the vertical ascents.
\vspace*{-2mm}

 \begin{center} \rule[0mm]{7.cm}{0.1mm} \end{center}
\vspace*{-2mm}

\noindent
{\bf{{Appendix~B}. Historical notes on the application of the third law}.}
\label{sec:third-law}

The dependence of $\theta_s$ on the absolute entropy, through the factor $\Lambda$ in Eq.~(\ref{eq:thetas}), arises because from the need to characterize a multi-component system whose relative composition (in our case between dry air and water vapor) is allowed to vary.

The recognition that the absolute value of the entropy are important for reacting, or multi-component systems, dates to \cite{Chatelier_1888} who first described the need to know the absolute values of entropy of reactants and products in order to be able to predict the stability of all chemical processes. 
Then, \cite{Nernst_1906} derived his ``theorem of heat'' but it is \cite{Planck1914,Planck1917} who really derived what is nowadays known as the Boltzmann equation $S = k \: \ln(W)$ with $k$ the Boltzmann constant. The absence of an additive constant corresponds to cancelling the entropy of all perfect crystalline state at zero Kelvin temperature (Third law of thermodynamics), due to the unique remaining number of configuration $W=1$ at \SI{0}{\kelvin}. 
\cite{Pauling1935} and \cite{Nagle1966} computed the residual entropy for ice at \SI{0}{\kelvin} ($\Delta S \approx \; $\SI{189}{\joule\per\kelvin\per\kg}), which must be taken into account for computing the entropy of water at any finite positive absolute temperature. The link between the third law of Planck and the principle of unattainability of absolute zero temperature derived by \cite{Nernst_1912} and studied by \cite{Simon_1927} has been recently clarified by \cite{Masanes_Oppenheim_2017}.

Values of absolute reference entropy of atmospheric gases (\ch{N2}, \ch{O2}, \ch{Ar}, \ch{H2O}, \ch{CO2}) used in \cite{HaufHoeller1987}, \cite{Marquet2011QJ} and \cite{stevenssiebesma2020} were already available in \cite{Kelley_1932}, \cite{Lewis_Randall_1961} and \cite{Gokcen_Reddy_1996}. They are now accurately determined and available in NIST-JANAF Tables \citep{Chase_1998}. The agreement between the various way to compute the absolute entropies can be fairly appreciated in 
Figs.~B1,
where the ``calorimetric'' and ``statistical-physics'' methods lead to the same results in the range of atmospheric temperatures up to better than \SI{\pm 0.6}{\percent} for \ch{H2O} and \ch{N2} and better than \SI{\pm 0.1}{\percent} for \ch{O2}, Ar and \ch{CO2}. The accuracy of the NIST-JANAF tables are indicated as being better than one tenth of the differences between calorimetric and the statistical methods.

It can be recalled that, if the third law is applied to \SI{0}{\kelvin}, the consequences of this hypothesis impact the atmospheric temperatures domain via the calorimetric method and the integrations made between \SI{0}{\kelvin} and any temperature $T$. 
The same is true for the statistical physics method, where the partition function $Z$ is computed with the hypothesis $S=k\:\ln(W)$ with no additive constant and with $S=0$ because $W=1$ at \SI{0}{\kelvin}.

The impacts of the hypotheses i) and ii) made at the end of section~\ref{sec:preliminaries}.\ref{sec:preliminaries_moist_air} concerning the constancy of the specific heats and the deviations from the ideal gas aspects remain small when compared to the data computed by the IAPWS and TEOS10 software (not shown).
Moreover, the absolute values of the entropies can easily be computed with TEOS10 if one takes into account the data from the thermodynamic tables \citep{Lewis_Randall_1961,Chase_1998}, or at least the liquid-water and dry-air absolute entropies given by \citet{Millero_1983} and \citet{Lemmon_al_2000}, respectively.

{\small
\bibliographystyle{ametsoc2014}
\bibliography{vartheta_arXiv_R2}
}

\end{document}